\documentclass{aa}

\usepackage{txfonts}
\usepackage{graphicx}
\usepackage{array}
\usepackage{booktabs}
\usepackage{amssymb}
\usepackage{subcaption}
\usepackage[colorlinks=true,allcolors=blue]{hyperref}

\def\Lsun{L_\odot}
\def\Msun{M_\odot}
\def\MJyr{\MJ\,\mathrm{yr}^{-1}}
\def\MMJyr{\MJ^2\,\mathrm{yr}^{-1}}

\def\Msunyr{M_\odot\,\mathrm{yr}^{-1}}
\def\flux{W\,m$^{-2}$~$\mu$m$^{-1}$\xspace}
\def\opacity{cm$^2$~g$^{-1}$\xspace}

\def\MJ{\ensuremath{M_\mathrm{J}}\xspace}
\def\RJ{\ensuremath{R_\mathrm{J}}\xspace}
\def\Mdot{\ensuremath{\dot{M}}\xspace}
\def\Mdotmin{\ensuremath{\dot{M}_\mathrm{min}}\xspace}
\def\Mp{\ensuremath{M_\mathrm{p}}\xspace}
\def\Rp{\ensuremath{R_\mathrm{p}}\xspace}
\def\DR{\ensuremath{\Delta R}\xspace}
\def\Rphot{\ensuremath{R_\mathrm{phot}}\xspace}
\def\Rin{\ensuremath{R_\mathrm{in}}\xspace}

\def\Lint{\ensuremath{L_\mathrm{int}}\xspace}
\def\LCPD{\ensuremath{L_\mathrm{CPD}}\xspace}
\def\LSED{\ensuremath{L_\mathrm{SED}}\xspace}
\def\Ltot{\ensuremath{L_\mathrm{tot}}\xspace}
\def\Lacc{\ensuremath{L_\mathrm{acc}}\xspace}
\def\Teff{\ensuremath{T_\mathrm{eff}}\xspace}
\def\TeffL{\ensuremath{T_\mathrm{eff,\,loc}}\xspace}
\def\Tdest{\ensuremath{T_\mathrm{dest}}\xspace}

\def\RHill{\ensuremath{R_\mathrm{Hill}}\xspace}
\def\Racc{\ensuremath{R_\mathrm{acc}}\xspace}
\def\vff{\ensuremath{v_\mathrm{ff}}\xspace}

\def\tauR{\ensuremath{\tau_\mathrm{R}}\xspace}
\def\kapR{\ensuremath{\kappa_\mathrm{R}}\xspace}
\def\kapRm{\ensuremath{\kappa_{\mathrm{R},\,\bullet}}\xspace}
\def\fpg{\ensuremath{f_\mathrm{d/g}}\xspace}

\def\ff{\ensuremath{f_\mathrm{fill}}\xspace}
\def\rp{\ensuremath{r_\mathrm{p}}\xspace}

\newcommand{\ffdeg}{\mbox{\ensuremath{.\!\!\degr}}}
\newcommand{\ffarcs}{\mbox{\ensuremath{.\!\!^{\prime\prime}}}}

\newcolumntype{L}[1]{>{\raggedright\let\newline\\\arraybackslash\hspace{0pt}}m{#1}}
\newcolumntype{C}[1]{>{\centering\let\newline\\\arraybackslash\hspace{0pt}}m{#1}}

\begin{document}

\title{MIRACLES: atmospheric characterization of directly imaged planets and substellar companions at 4--5~$\mu$m\thanks{Based on observations collected at the European Southern Observatory, Chile, ESO No. 095.C-0298(A), 097.C-0206(A), 1100.C-0481(D), and 0102.C-0649(A).}}

\subtitle{II. Constraints on the mass and radius of the enshrouded planet PDS~70~b}

\titlerunning{MIRACLES II. Constraints on the mass and radius of the enshrouded planet PDS~70~b}

\author{
T.~Stolker\inst{\ref{eth}}
\and G.-D.~Marleau\inst{\ref{tueb},\ref{bern},\ref{mpia}}
\and G.~Cugno\inst{\ref{eth}}
\and P.~Molli\`{e}re\inst{\ref{mpia}}
\and S.\,P.~Quanz\inst{\ref{eth}}
\and K.\,O.~Todorov\inst{\ref{api}}
\and J.~K\"{u}hn\inst{\ref{bern}}
}

\institute{
Institute for Particle Physics and Astrophysics, ETH Zurich, Wolfgang-Pauli-Strasse 27, 8093 Zurich, Switzerland\\
\email{tomas.stolker@phys.ethz.ch}
\label{eth}
\and Institut f\"ur Astronomie und Astrophysik, Universit\"at T\"{u}bingen, Auf der Morgenstelle 10, 72076 T\"{u}bingen, Germany
\label{tueb}
\and Physikalisches Institut, Universit\"at Bern, Gesellschaftsstrasse 6, 3012 Bern, Switzerland
\label{bern}
\and Max-Planck-Institut f\"ur Astronomie, K\"{o}nigstuhl 17, 69117 Heidelberg, Germany
\label{mpia}
\and Anton Pannekoek Institute for Astronomy, University of Amsterdam, Science Park 904, 1090 GE Amsterdam, The Netherlands
\label{api}
}

\date{Received ?; accepted ?}

\abstract
{The circumstellar disk of PDS 70 hosts two forming planets, which are actively accreting gas from their environment. The physical and chemical characteristics of these planets remain ambiguous due to their unusual spectral appearance compared to more evolved objects. In this work, we report the first detection of PDS~70~b in the Br$\alpha$ and $M'$ filters with VLT/NACO, a tentative detection of PDS~70~c in Br$\alpha$, and a reanalysis of archival NACO $L'$ and SPHERE $H23$ and $K12$ imaging data. The near side of the disk is also resolved with the Br$\alpha$ and $M'$ filters, indicating that scattered light is non-negligible at these wavelengths. The spectral energy distribution (SED) of PDS~70~b is well described by blackbody emission, for which we constrain the photospheric temperature and photospheric radius to $\Teff=1193 \pm 20$~K and $R=3.0 \pm 0.2$~$\RJ$. The relatively low bolometric luminosity, $\log(L/L_\odot) = -3.79 \pm 0.02$, in combination with the large radius, is not compatible with standard structure models of fully convective objects. With predictions from such models, and adopting a recent estimate of the accretion rate, we derive a planetary mass and radius in the range of $\Mp\approx0.5$--1.5~\MJ and $\Rp\approx1$--2.5~\RJ, independently of the age and post-formation entropy of the planet. The blackbody emission, large photospheric radius, and the discrepancy between the photospheric and planetary radius suggests that infrared observations probe an extended, dusty environment around the planet, which obscures the view on its molecular composition. Therefore, the SED is expected to trace the reprocessed radiation from the interior of the planet and/or partially from the accretion shock. The photospheric radius lies deep within the Hill sphere of the planet, which implies that PDS~70~b not only accretes gas but is also continuously replenished by dust. Finally, we derive a rough upper limit on the temperature and radius of potential excess emission from a circumplanetary disk, $\Teff\lesssim256$~K and $R\lesssim245~\RJ$, but we do find weak evidence that the current data favors a model with a single blackbody component.}


\maketitle

\section{Introduction}
\label{sec:introduction}

The formation of planets occurs in circumstellar disks (CSDs) around pre-main sequence stars. Spatially resolved observations have revealed a ubiquity of substructures in the gas and dust distribution of those disks, such as gaps and spiral arms \citep[e.g.,][]{andrews2018,avenhaus2018}. These features may point to the gravitational interaction of embedded planets with their natal environment \citep[e.g.,][]{pinilla2012b,dong2015}, but the direct detection of these potential protoplanets remains challenging \citep[e.g.,][]{currie2019,cugno2019}, possibly due to their low intrinsic brightness and extinction effects by dust \citep[e.g.,][]{brittain2020,sanchis2020}. Nevertheless, direct detections of forming planets are critical to advance our empirical understanding of the physical processes by which planets accumulate gas and dust from, and interact with, their circumstellar environment.

The CSD of PDS 70 is a unique example in which two embedded planets were directly detected with high-resolution instruments. PDS~70 is a weak-line T~Tauri, K7-type \citep{pecaut2016} star with an estimated age of $5.4 \pm 1.0$~Myr \citep{mueller2018}; it is surrounded by a gapped accretion disk \citep{hashimoto2012} and is located in the Scorpius-Centaurus OB association \citep{gregorio2002,preibisch2008}. \citet{keppler2018} discovered PDS~70~b within the gap of the disk with SPHERE and archival $L'$ band data. This planet is located at a favorable position (close to the major axis of the disk) where projection and extinction effects are minimized. Later, a second planetary companion, PDS~70~c, was discovered by \citet{haffert2019} in H$\alpha$, together with a detection of H$\alpha$ emission from PDS~70~b \citep[see also][]{wagner2018}. Such measurements of hydrogen emission lines place constraints on the physics of the accretion flow and shock, extinction, and the mass and accretion rate of the planets \citep{thanathibodee2019,aoyama2019,hashimoto2020}.

While PDS~70~b and~c have been suggested to be planetary-mass objects and have been confirmed to be comoving with PDS~70 \citep{keppler2018,mueller2018,haffert2019,mesa2019}, their atmospheric and circumplanetary characteristics remain poorly understood. The $H$ and $K$ band fluxes of PDS~70~b are consistent with a mid L-type object, but its near-infrared (NIR) colors are redder than those of field dwarfs and low-gravity objects \citep{keppler2018}. A comparison with cloudy atmosphere models by \citet{mueller2018} shows that a wide range of temperatures (1000--1600~K) and radii (1.4--3.7~$\RJ$) could describe the spectral energy distribution (SED) from the $Y$ to $L'$ bands. A detailed SED analysis by \citet{christiaens2019} has revealed an excess of the $K$ band emission with respect to predictions by atmospheric models. The authors show that the SED is consistent with a combination of emission from a planet atmosphere (1500--1600~K) and a circumplanetary disk (CPD). Most recently, \citet{wang2020} presented NIRC2 $L'$ imaging and analyzed the SED with atmospheric models and blackbody spectra. From this, the authors conclude that the data is best described by a blackbody spectrum with $\Teff = 1204^{+52}_{-53}$~K and $R = 2.72^{+0.39}_{-0.34}~\RJ$.

In this work, we report the first detection of PDS~70~b in the 4--5~$\mu$m range as part of the MIRACLES survey \citep{stolker2020}. The object was observed with NACO at the Very Large Telescope (VLT) in Chile and detected with both the Br$\alpha$ and $M'$ filters. Mid-infrared (MIR) wavelengths are in particular critical to uncover potential emission from a circumplanetary environment. We will analyze the photometry, colors, and SED of the object to gain insight into its main characteristics.

\section{Observations and data reduction}
\label{sec:observations}

\subsection{High-contrast imaging with VLT/NACO}
\label{sec:imaging}

We observed PDS 70 with VLT/NACO \citep{lenzen2003,rousset2003} in the NB4.05 (Br$\alpha$; $\lambda_0 = 4.05$~$\mu$m, $\Delta\lambda = 0.02$~$\mu$m) and $M'$ ($\lambda_0 = 4.78$~$\mu$m, $\Delta\lambda = 0.59$~$\mu$m) filters (ESO program ID: 0102.C-0649(A)) as part of the MIRACLES survey, which aims at the systematic characterization of directly imaged planets and brown dwarfs at 4--5~$\mu$m \citep{stolker2020}. The data were obtained without coronagraph in pupil-stabilized mode, while dithering the star across the detector to sample the thermal background emission. The total telescope time was 3~hours and 4~hours for the NB4.05 and $M'$ filters, respectively, split over multiple nights with observing blocks (OBs) of 1 hour each. This resulted in a total of 1.75 and 2~hours of on-source telescope time for NB4.05 and $M'$. A detailed description of the observing strategy for the MIRACLES survey is available in \citet{stolker2020} but a few specifics for the observations of PDS~70 are provided here.

The observations with the NB4.05 filter were executed on UT 2019 February 23 and UT 2019 March 15. A detector integration time (DIT) of 1.0~s and NDIT of 61 or 65 was used, resulting in 1680 (first and second OB) and 1792 (third OB) frames. During the first night, two OBs were executed in good conditions (seeing $\lesssim$0\ffarcs8) while during the second night (i.e., the third OB), the conditions were slightly worse (0\ffarcs75--0\ffarcs95), resulting in an average angular resolution of 115~mas (1~FWHM). Aperture photometry (2~FWHM in diameter) of the star revealed flux variations of 4.6\% across the three datasets, which in particular reflects the variable conditions during the third OB. The total, non-intermittent, and non-overlapping field rotation was 50~deg but gaps in the parallactic angle range between OBs helped with minimizing the self-subtraction during post-processing.

With a similar setup, we observed the target with the $M'$ filter on UT 2019 February 20, 21, and 22, with two OBs executed during the second night. The detector was windowed to a field of view of $256 \times 256$~pixels to allow for a short integration time of 35~ms without frame loss. With an NDIT of 1500 integrations and 14 exposures (i.e., data cubes) for each of the two dithering positions, this resulted in 42000 frames per OB. The seeing was approximately stable during three of the observations with average values in the range of 0\ffarcs7--0\ffarcs8. During the second OB, the seeing was about 1\ffarcs0--1\ffarcs2 with a short increase to 2\ffarcs0. After a frame selection and combining the data from the four OBs, the stellar flux varied by about 6.5\% and the FWHM of the PSF was 134~mas. The total, continuous field rotation was 56~deg, but 82~deg if the gaps in the parallactic angle coverage are included.

\subsection{Data processing and calibration}
\label{sec:data_processing}

The data were processed with \texttt{PynPoint}\footnote{\url{https://pynpoint.readthedocs.io}} which is a generic, end-to-end pipeline for high-contrast imaging data \citep{amara2012,stolker2019}. We used the latest release of the package (version 0.8.3) for the pre- and post-processing, and the relative photometric and astrometric calibration. The pre-processing was done for each dataset separately and the frames from the different OBs were combined before the PSF subtraction. We used an implementation of full-frame principal component analysis \citep[PCA;][]{amara2012,soummer2012} to remove the quasi-static structures of the stellar PSF.

In general, we followed the processing and calibration procedure that is described in \citet{stolker2020}. However, in addition to subtracting the mean background (based on the adjacent data cubes in which the star was dithered to a different detector position) we also applied an additional correction with PCA \citep{hunziker2018}. Specifically, we decomposed the stack of all background images (after subtracting the mean of the stack) at a given dithering position and projected the science data on the first principal component (PC). The central region (8~FWHM in diameter) was masked during the projection but included when subtracting the model. This provided better results on visual inspection compared to a mean background subtraction alone.

\begin{figure*}
\centering
\includegraphics[width=0.9\linewidth]{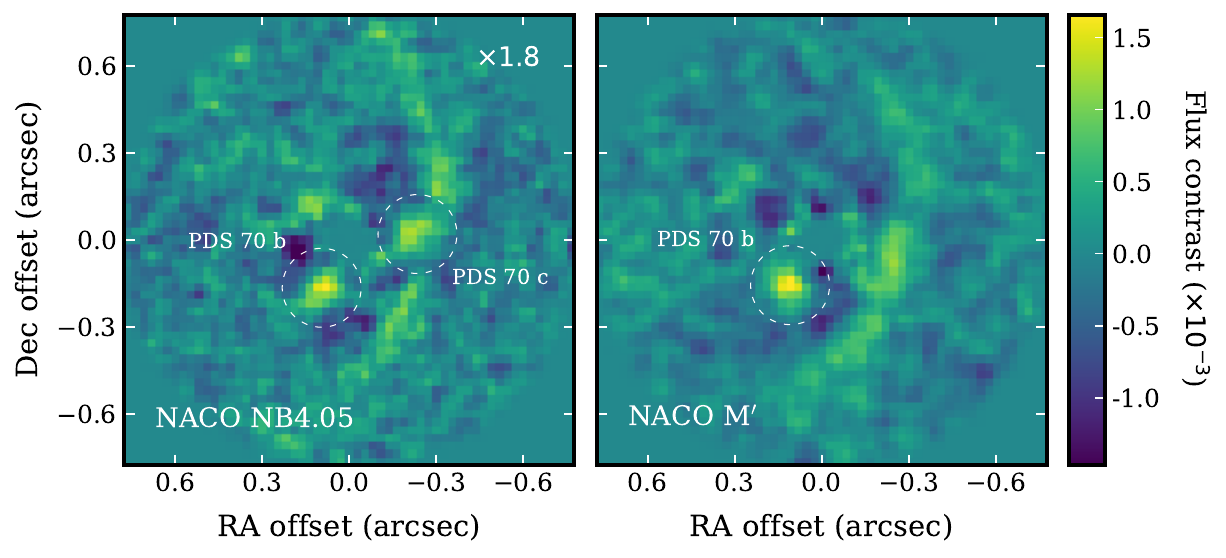}
\caption{Detection of the PDS 70 planetary system and CSD with the NACO NB4.05 (\emph{left panel}) and $M'$ (\emph{right panel}) filters. The images show the mean-combined residuals of the PSF subtraction on a color scale that has been normalized to the peak of the stellar PSF. The flux in the NB4.05 image has been increased by a factor of 1.8 for clarity. The detected emission from PDS~70~b and~c (only marginal in NB4.05) is encircled. North is up and east is left.}
\label{fig:images}
\end{figure*}

After pre-proccessing and combining the OBs, subsets of 8 (NB4.05) and 330 ($M'$) images were mean-collapsed, resulting in a final stack of 501 and 502 images for NB4.05 and $M'$, respectively. Then, we extracted the photometry and astrometry of the companions relative to their star in the following way. First, the dependence on the number of PCs was tested (1--5~PCs for NB4.05 and 1--10~PCs for $M'$), second, we used an MCMC approach to estimate the statistical uncertainty for a fixed number of PCs by removing the planet signal with a negative copy of the PSF, and thirdly, a bias and systematic uncertainty was estimated by injecting and retrieving artificial planets (see \citealp{stolker2020} for details). For the calibration, we used a field of view of 57~pixels, we subtracted three (NB4.05) and five ($M'$) PCs, and we applied an one-on-one injection of the PSF templates (the stellar flux had remained within the linear regime of the detector). The estimation of a potential bias and systematic error is challenging since the planet is only at 1.5~$\lambda$/D in $M'$, and both disk signal and noise residuals are present at the same separation (see Fig.~\ref{fig:images}). Therefore, to not introduce a bias, we excluded position angles with relatively bright disk or noise features for the estimation of the systematic error (see Table~\ref{table:photometry}).

From the relative calibration, we determined the apparent magnitudes in the NB4.05 and $M'$ filters. We first used the \texttt{species}\footnote{\url{https://species.readthedocs.io}} toolkit \citep{stolker2020} to convert the 2MASS $JHK$, and WISE $W1$ and $W2$ magnitudes of the PDS 70 system into fluxes. We then fitted a power law function to these in log-log space. The stellar magnitudes in NB4.05 and $M'$ were then computed by integrating the model spectrum across the filter profiles (see Table~\ref{table:photometry}). We note that this approach assumes that the photometry in the considered spectral range is dominated by continuum emission from the star and inner disk. Therefore, potential Br$\alpha$ emission due to accretion onto the star is ignored, but that is a reasonable assumption given the low accretion rate of $(0.6$--$2.2) \times 10^{-10}~\Msunyr$ for PDS~70 \citep{thanathibodee2020}.

\subsection{Reanalysis of archival data}
\label{sec:archival_data}

In addition to the new NB4.05 and $M'$ data, we reanalyzed archival NACO $L'$ data from \citet{keppler2018} (ESO program ID: 097.C-0206(A)), in line with the systematic 3--5~$\mu$m analysis for the MIRACLES survey, and additionally coronagraphic SPHERE $H23$ and $K12$ data from \citet{keppler2018} and \citet{mueller2018} (ESO program IDs: 095.C-0298(A) and 1100.C-0481(D)). Below, we provide a few details on the data quality and processing, but we refer to the respective papers for more information on these datasets. The calibration was done in a similar way as the NB4.05 and $M'$ data. While the $H$ band is also covered by the NIR spectrum that we adopted from \citet{mueller2018} (see Sect.~\ref{sec:modelling_approach}), the $K$ band flux was in particular critical for estimating the photospheric temperature and radius of PDS~70~b (see Sect.~\ref{sec:sed_analysis}).

The NACO $L'$ data were obtained with seeing conditions in the range of 0\ffarcs55--0\ffarcs7. We removed 18\% of the frames (based on aperture photometry at the position of the star) after which the stellar flux varied by 29\% across the dataset. The flux had not saturated the detector so we applied an one-on-one PSF injection during the relative calibration. Therefore, the variation in the stellar flux is not expected to have introduced a bias in the extracted planet flux. A total of 14464 frames were selected across a parallactic rotation of 85~deg.

The archival $H23$ and $K12$ datasets had been obtained with the IR dual-band imager \citep[IRDIS;][]{dohlen2008,vigan2010} of SPHERE \citep{beuzit2019}. We analyzed both $H23$ epochs from \citet{keppler2018} but only use the results from UT 2015 May 04 since the second dataset (UT 2015 June 01) was obtained in poor observing conditions with a seeing larger than 1\arcsec. During the first epoch, the seeing was 0\ffarcs35 at the start of the observations, but degraded to $>$1\arcsec at 1/3 of the sequence. The stellar halo appeared bright and asymmetric, possibly due to a low-wind effect ($\sim$5~m~s$^{-1}$) and/or a wind-driven halo \citep{cantalloube2018}. We only used 30 frames that were obtained in good conditions, which were selected by measuring the flux of the background star at $\sim$2\ffarcs4 north of PDS~70. Similarly, we only used the off-axis PSF exposures from the start of the observations because these were obtained in conditions that were similar to the selected frames with the star behind the coronagraph. The flux in the unsaturated PSF exposures has been scaled to the coronagraphic data to account for the difference in exposure time and the transmission of the neutral density filter.

\begin{table*}
\caption{Photometry and error budget.}
\label{table:photometry}
\centering
\bgroup
\def\arraystretch{1.25}
\begin{tabular}{L{2.2cm} C{1.55cm} C{1.65cm} C{0.85cm} C{1.55cm} C{0.85cm} C{1.65cm} C{1.65cm} C{2.35cm}}
\hline\hline
Filter & MCMC contrast & Bias offset & Calib. error & Final contrast & Star & Apparent magnitude & Absolute magnitude & Flux \\
& (mag) & (mag) & (mag) & (mag) & (mag) & (mag) & (mag) & (W m$^{-2}$ $\mu$m$^{-1}$) \\
\hline
\multicolumn{9}{l}{\emph{PDS 70 b}}\\
SPHERE $H2$ & $9.11 \pm 0.11$ & $0.02 \pm 0.18$ & 0.03 & $9.13 \pm 0.21$ & $8.99$ & $18.12 \pm 0.21$ & $12.85 \pm 0.21$ & $7.41(1.42) \cdot 10^{-17}$ \\
SPHERE $H3$ & $9.03 \pm 0.11$ & $0.02 \pm 0.14$ & 0.03 & $9.05 \pm 0.18$ & $8.92$ & $17.97 \pm 0.18$ & $12.70 \pm 0.18$ & $7.20(1.21) \cdot 10^{-17}$ \\
SPHERE $K1$ & $8.09 \pm 0.03$ & $0.00 \pm 0.03$ & 0.01 & $8.09 \pm 0.04$ & $8.57$ & $16.66 \pm 0.04$ & $11.39 \pm 0.04$ & $1.05(0.04) \cdot 10^{-16}$ \\
SPHERE $K2$ & $7.90 \pm 0.04$ & $0.00 \pm 0.04$ & 0.01 & $7.90 \pm 0.06$ & $8.47$ & $16.37 \pm 0.06$ & $11.09 \pm 0.06$ & $1.06(0.06) \cdot 10^{-16}$ \\
NACO $L'$ & $6.77 \pm 0.19$ & $0.03 \pm 0.14$ & --- & $6.80 \pm 0.24$ & $7.86$ & $14.66 \pm 0.24$ & $9.39 \pm 0.24$ & $7.21(1.62) \cdot 10^{-17}$ \\
NACO NB4.05 & $6.90 \pm 0.23$ & $0.01 \pm 0.14$ & --- & $6.91 \pm 0.27$ & $7.77$ & $14.68 \pm 0.27$ & $9.40 \pm 0.27$ & $5.35(1.36) \cdot 10^{-17}$ \\
NACO $M'$ & $6.12 \pm 0.19$ & $0.03 \pm 0.19$ & --- & $6.15 \pm 0.27$ & $7.65$ & $13.80 \pm 0.27$ & $8.52 \pm 0.27$ & $6.56(1.63) \cdot 10^{-17}$ \\
\hline
\multicolumn{9}{l}{\emph{PDS 70 c}}\\
NACO NB4.05 & $7.06 \pm 0.21$ & $0.11 \pm 0.09$ & --- & $7.17 \pm 0.23$ & $7.77$ & $14.94 \pm 0.23$ & $9.67 \pm 0.23$ & $4.19(0.89) \cdot 10^{-17}$ \\
\hline
\end{tabular}
\egroup
\end{table*}

There are two archival SPHERE/IRDIS $K12$ datasets available, which had been obtained on UT 2016 May 15 and 2018 February 25. We analyzed both datasets but only used the results from the second epoch because the assessment of the first epoch revealed large-scale noise residuals after the PSF subtraction, which may have biased the photometry. The second dataset was obtained in good observing conditions but the seeing degraded toward the end of the sequence. Therefore, similar to the $H23$ data, we selected 24 frames from the start of the sequence, based on the photometry of the background star, and the unsaturated exposures from the start of the observations.

\section{Results}
\label{sec:results}

\subsection{Detection of the PDS~70 system}
\label{sec:detection}

The mean-combined residuals from the PSF subtraction after subtracting two (NB4.05) and three ($M'$) PCs are presented in Fig.~\ref{fig:images}. The choice of the number of PCs for the image is dictated by the brightness of the planets; to characterize them a somewhat larger number of PCs was removed (see Sect.~\ref{sec:data_processing}) to better suppress the residual speckle noise. The images reveal a bright source at the expected position of PDS~70~b.

While planet~b is visible in both filters, planet~c is only marginally detected with the NB4.05 filter and not visible in the $M'$ image. Here, the position of planet~c relative to the near side of the disk may have prevented a detection in $M'$ due to the reduced angular resolution compared to NIR wavelengths. In the NB4.05 image, planet~c is blended with the disk signal, and therefore the extracted flux is potentially biased. We estimated the bias due to the CSD signal by injecting and retrieving the contrast of an artificial planet at a location with comparable disk flux but somewhat offset from the c planet, yielding an approximate correction of $\sim$0.1~mag (see Table~\ref{table:photometry}).

The results from the photometric extraction of the companions are listed in Table~\ref{table:photometry}, both for the new and archival data. The final contrast is calculated by adding the bias offset and combining the error components in quadrature. The error budget of the planet photometry is dominated by the error from the relative calibration while the error on the stellar magnitude (expected to be a few tens of a magnitude) is negligible. For the coronagraphic SPHERE $H23$ and $K12$ data, we have included an additional error component that was derived from the flux of the background star, which varied by about $\sim$1\% after the frame selection. The astrometry is available in \ref{table:astrometry} of Appendix~\ref{sec:astrometry} but these results will not be analyzed.

In addition to the point sources, also scattered light from the near side of the gap edge of the CSD, which is illuminated by the central star, is visible in both datasets. Therefore, the scattering opacity of the dust grains in the disk surface is non-negligible even at these relatively long wavelengths. Interestingly, only the near side of the disk is visible which points to an asymmetry in the scattering phase function of the dust. This finding suggests that the dust grains are comparable to or larger than the observed wavelength (4--5~$\mu$m).

\subsection{Color and magnitude comparison}
\label{sec:color_magnitude}

\subsubsection{Color--magnitude diagram}
\label{sec:color_mag_diagrm}

\begin{figure*}
\centering
\includegraphics[width=0.75\linewidth]{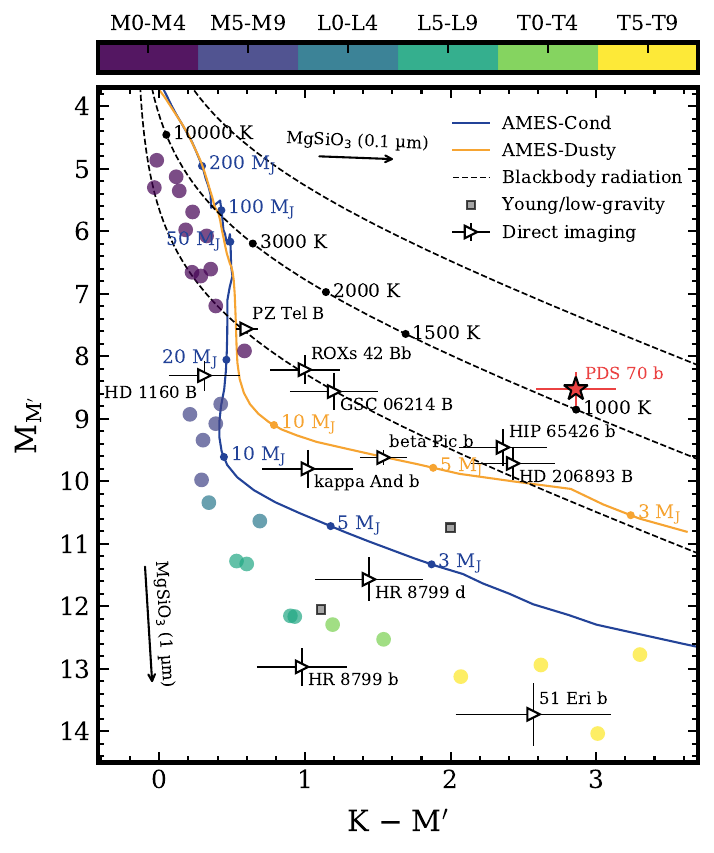}
\caption{Color--magnitude diagram of M$_{M'}$ versus $K$~--~$M'$. The field objects are color-coded by M, L, and T spectral types (see discrete colorbar), the young and low-gravity objects are indicated with a \emph{gray square}, and the directly imaged companions are labeled individually. PDS~70~b is highlighted with a \emph{red star}. The \emph{blue} and \emph{orange} lines show the synthetic colors computed from the AMES-Cond and AMES-Dusty evolutionary tracks at an age of 5~Myr. Blackbody radiation curves are shown for 8~$\RJ$, 4~$\RJ$, and 2~$\RJ$ (\emph{black dashed lines}, from top to bottom). The \emph{black arrows} indicate the reddening by MgSiO$_3$ grains with a mean radius of 0.1 and 1~$\mu$m, and $A_{M'}$ of 0.05 and 2~mag, respectively.}
\label{fig:color_magnitude}
\end{figure*}

The absolute brightness of PDS~70~b in the new and archival data is derived from the calibrated magnitudes in Table~\ref{table:photometry} and the \emph{Gaia} distance of $113.4 \pm 0.5$~pc \citep{gaia2018}. We determined absolute magnitudes of $9.40 \pm 0.27$~mag and $8.52 \pm 0.27$~mag in the NB4.05 and $M'$ filters, respectively. The uncertainty on the parallax is negligible in the error budget. With the $K1$ and $M'$ magnitudes, we place PDS~70~b in a color--magnitude diagram to show its photometric characteristics with respect to field and low-gravity dwarfs \citep{dupuy2012,dupuy2013,liu2016}, other directly imaged planets planets and brown dwarfs \citep{marois2010,bonnefoy2011,ireland2011,galicher2011,bailey2013,bonnefoy2014,daemgen2017,chauvin2017,delorme2017,rajan2017,stolker2019,stolker2020}, predictions by the AMES-Cond and AMES-Dusty evolutionary and atmospheric models \citep{chabrier2000,allard2001,baraffe2003}, and blackbody spectra.  The color--magnitude diagram was created with the \texttt{species} toolkit \citep{stolker2020} and is shown in Fig.~\ref{fig:color_magnitude}. We note that the SPHERE $K1$ magnitude was adopted for PDS~70~b, HIP~65426~b, and HD~206893~B. Since the $K1$ filter is close to the central wavelength of a typical $K$ band filter, the color between such filters is $\lesssim$0.1~mag, which has been quantified by considering all available DRIFT-PHOENIX spectra \citep{helling2008} with $\Teff$ in the range of 1000--2000~K. Such a color effect is small compared to the uncertainty on the $K$~--~$M'$ color of these three objects.

The $M'$ flux of PDS~70~b is consistent with a mid to late M-type field dwarf and comparable in brightness to ROXs~42~Bb and GSC~06214~B, which are both young, planetary-mass companions. The latter is known to have a circumsubstellar disk \citep{bowler2011}. Compared to the L-type directly imaged planets $\beta$~Pic~b and HIP~65426~b, PDS~70~b is brighter in $M'$ by $\sim$1~mag. In addition to the absolute brightness, we derived a $K1$~--~$M'$ color of $2.86 \pm 0.27$~mag, which is significantly redder than any of the planets and brown dwarfs in the color--magnitude diagram. Specifically, PDS~70~b is about 2~mag redder than the young, planetary-mass companions and 1.3~mag redder than $\beta$~Pic~b. Most comparable in color are HIP~65426~b and HD~206893~B but the difference is still 0.4~mag and these objects are $\gtrsim$1~mag fainter in $M'$. Interestingly, these are two of the reddest low-mass companions \citep{milli2017,cheetham2019}, with unusual $M'$ colors that might be caused by enhanced cloud densities close to their photosphere \citep{stolker2020}.

The empirical comparison shows that PDS~70~b is brighter and/or redder than any of the other directly imaged planets. In addition, we compare the data with synthetic photometry from the AMES-Cond (cloudless) and AMES-Dusty (efficient mixing of dust grains) models, which have been computed from the isochrone data at an age of 5~Myr. The comparison in Fig.~\ref{fig:color_magnitude} shows that the observed flux in $M'$ is about 1.6~mag brighter than the AMES-Dusty predictions for an object of the same color, which would have a mass of 3--4~$\MJ$. This flux difference corresponds to a factor of $\approx$2.1 in radius. In the model spectra, the dust causes a veiling of the molecular features and a shift of the photosphere to higher (cooler) altitudes. Consequently, the IR colors become redder and the $M'$ flux larger, in particular because of weaker CO absorption at 4.6~$\mu$m. While the radius had been calculated self-consistently in these models, the offset with the PDS~70~b magnitude may indicate that either the radius is larger than predicted and/or the atmosphere is even dustier than what is modeled. 

A comparison of the photometric characteristics with the synthetic fluxes from a blackbody spectrum shows indeed that PDS~70~b is consistent with a blackbody temperature of $\sim$1000~K and a radius of $\approx$5~$\RJ$ (see Sect.~\ref{fig:sed} for a more detailed estimation of the blackbody parameters). This is in tension with the predicted radii in the AMES-Dusty and AMES-Cond models, either of which have $\approx1.4$--1.8~$\RJ$ for 1--10~$\MJ$ at 5~Myr (see isochrones in Fig.~\ref{fig:luminosity}). As was pointed out some time ago \citep{fortney2005,marley2007}, at these early ages ($\lesssim$50--100~Myr) the (arbitrary) choice of the starting luminosity or radius in the models still matters a lot; put differently, the planet may have formed (much) later than the star. Whether considering a younger cooling age sufficiently alleviates the tension is discussed in Sect.~\ref{sec:sed_constraints}.

PDS~70~b is located in the gap of a CSD and is actively accreting from its environment. Therefore, the planet might be partially obscured by (dusty) material in its vicinity, which is expected to attenuate the planet's spectrum. To understand the impact of the dust on the color and magnitude of the object, we show reddening vectors in Fig.~\ref{fig:color_magnitude} for spherical grains with a homogeneous, crystalline enstatite composition \citep{scott1996,jaeger1998}. The extinction cross sections were calculated with \texttt{PyMieScatt} \citep{sumlin2018} by assuming a log-normal size distribution with a geometric standard deviation of 2 \citep{ackerman2001}. For grains with a geometric mean radius of 0.1~$\mu$m, the extinction would cause a reddening of the $K$~--~$M'$ color, which would result in an under- and overestimated blackbody temperature and radius, respectively. For 1~$\mu$m grains, the color is close to gray so potential extinction would cause an underestimation of the planet radius. The radius of PDS~70~b will be estimated and discussed in more detail in Sect.~\ref{sec:sed_constraints}.

\subsubsection{Color--color diagram}
\label{sec:color_color_diagrm}

\begin{figure*}
\centering
\includegraphics[width=0.75\linewidth]{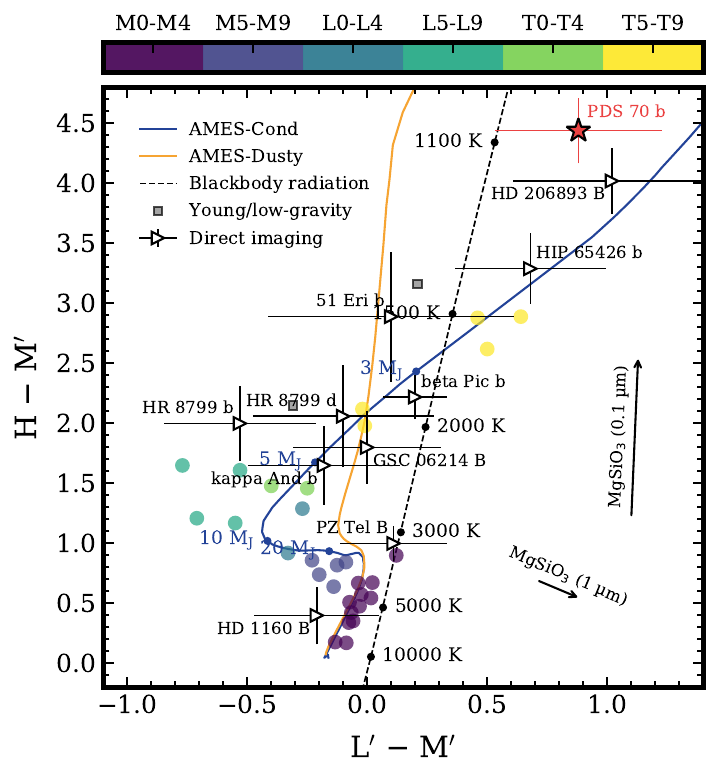}
\caption{Color--color diagram of $H$~--~$M'$ versus $L'$~--~$M'$. The field objects are color-coded by M, L, and T spectral types (see discrete colorbar), the young and low-gravity dwarf objects are indicated with a \emph{gray square}, and the directly imaged companions are labeled individually. PDS~70~b is highlighted with a \emph{red star}. The \emph{blue} and \emph{orange} lines show the synthetic colors computed from the AMES-Cond and AMES-Dusty evolutionary tracks at an age of 5~Myr. The \emph{black dashed line} shows the synthetic colors of a blackbody spectrum. The \emph{black arrows} indicate the reddening by MgSiO$_3$ grains with a mean radius of 0.1 and 1~$\mu$m, and $A_{M'}$ of 0.05 and 5~mag, respectively.}
\label{fig:color_color}
\end{figure*}

While color--magnitude diagrams reveal trends related to the intrinsic brightness of an object, color--color diagrams are independent of the distance and radius. Therefore, they are a useful diagnostic for understanding correlations between colors which are related to the atmospheric characteristics. In the case of a forming planet, the interpretation is more complicated because the colors are also affected by the accretion luminosity and the presence of circumplanetary material. This may cause a reddening of the IR fluxes due to reprocessed radiation and extinction of the atmospheric flux.

The data from Fig.~\ref{fig:color_magnitude} are used together with available $H$ (either broad- or narrowband) and $L'$ photometry of directly imaged companions \citep{biller2010,ireland2011,currie2012,currie2013,currie2014,bonnefoy2014,milli2017,chauvin2017,rajan2017,keppler2018}. We created a color--color diagram of $H$~--~$M'$ versus $L'$~--~$M'$ with \texttt{species}, which is displayed in Fig.~\ref{fig:color_color}. PDS~70~b is positioned in a red part of the diagram with a $H$~--~$M'$ color of $4.44 \pm 0.27$~mag and a $L'$~--~$M'$ color of $0.88 \pm 0.35$~mag. For $H$~--~$M'$, we computed the synthetic MKO $H$ band photometry ($18.24 \pm 0.04$~mag) from the SPHERE spectrum of \citet{mueller2018}, although the difference between the broadband $H$ and narrowband $H2$ photometry is only $\sim$0.1~mag. Both colors are consistent with HD~206893B and the $L'$~--~$M'$ color is also comparable to HIP~65426~b.

The color characteristics of PDS~70~b are clearly distinct from more evolved objects. Specifically, the sequence of field objects and cloudless atmosphere models show approximately gray colors at high temperatures, while toward lower temperatures the $L'$~--~$M'$ color becomes bluer and then redder because of CO and CH$_4$ absorption, respectively \citep[see e.g.,][]{stolker2020}. Similarly, the increasing strength of H$_2$O absorption in the $H$ band causes a redder $H$~--~$M'$ color toward lower temperatures. Interestingly, the $H$ band spectrum of PDS~70~b shows only weak evidence of H$_2$O absorption \citep{mueller2018} so the origin of the very red $H$~--~$M'$ color is presumably different.

Spectra of giant planets and brown dwarfs are usually not well described by blackbody emission due to molecular absorption which causes a strong variation in the photosphere temperature with wavelength. Indeed, the comparison of the colors in Fig.~\ref{fig:color_color} shows that, for a given temperature, the blackbody colors are redder than the colors of M- and L-type field objects, as well as the predictions from the atmospheric models. Several of the directly imaged objects lie close to the blackbody curve but the uncertainties (on the $M'$ flux in particular) are large. The spectrum of a low-gravity atmosphere may indeed approach a blackbody spectrum if the quasi-continuum cross-sections of the dust grains dominate the atmospheric opacity.

\subsection[The spectral energy distribution from 1 to 5 um]{The spectral energy distribution from 1 to 5~$\mu$m}
\label{sec:sed_analysis}

\subsubsection{Modeling approach of the SED}
\label{sec:modelling_approach}

\begin{figure*}
\centering
\begin{subfigure}{\textwidth}
\centering
\includegraphics[width=\linewidth]{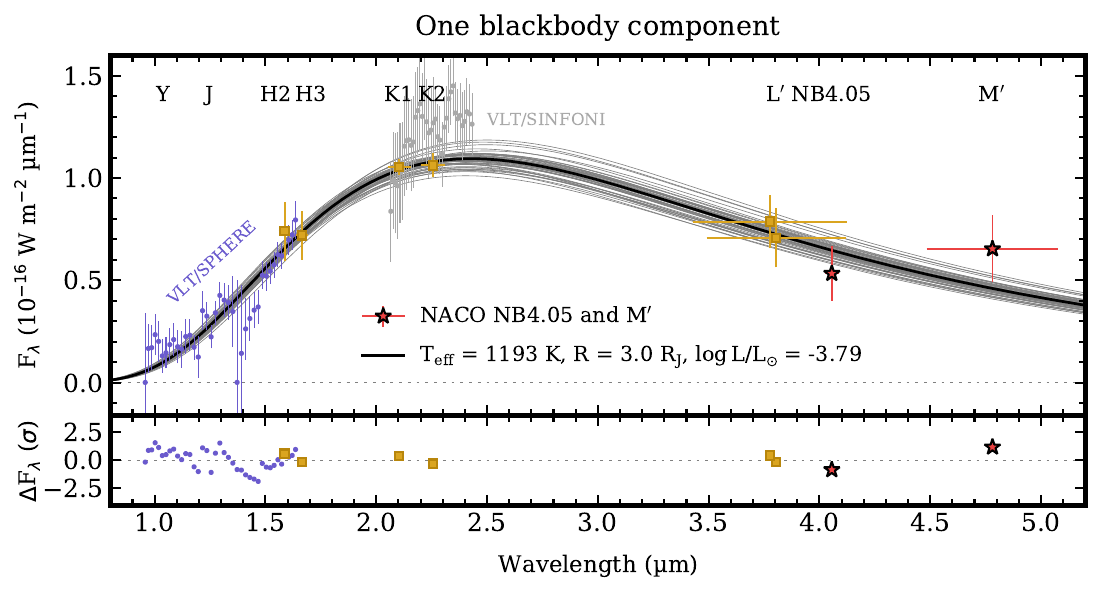}
\end{subfigure}
\begin{subfigure}{\textwidth}
\centering
\includegraphics[width=\linewidth]{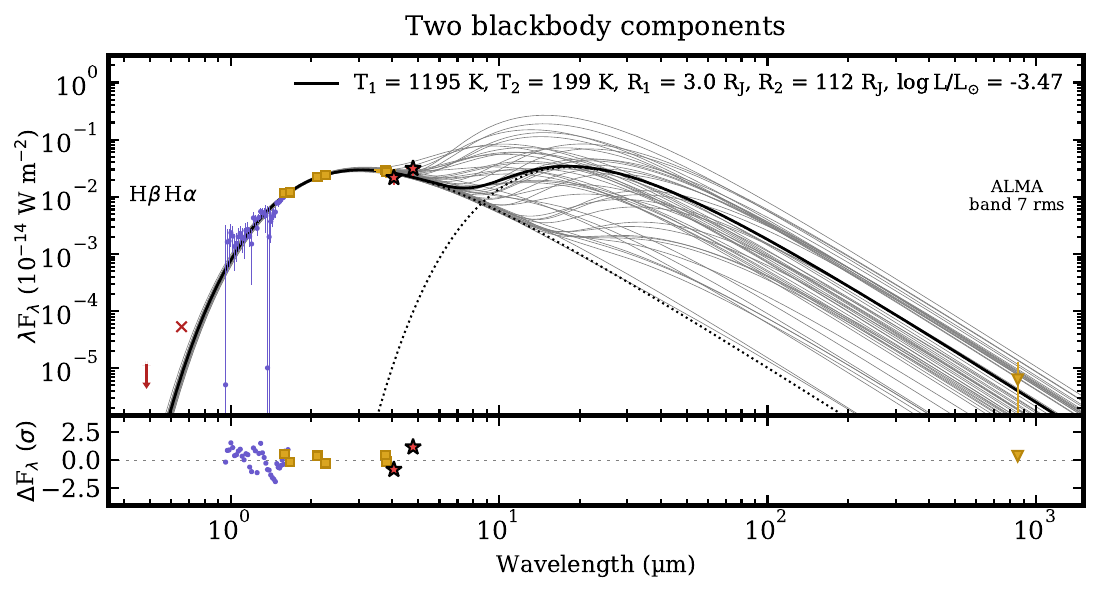}
\end{subfigure}
\caption{Spectral energy distribution of PDS~70~b. The \emph{top} and \emph{bottom panel} show the results from fitting one and two blackbody components, respectively (the flux units are different between the two panels). The photometric and spectroscopic data (\emph{colored markers}) are shown in comparison with the best-fit blackbody spectrum (\emph{black line}), and randomly drawn samples from the posterior distribution (\emph{gray lines}). The residuals are shown relative to the data uncertainties. The H$\alpha$ and H$\beta$ (upper limit) fluxes \citep{hashimoto2020} are shown for reference but were not used in the fit.}
\label{fig:sed}
\end{figure*}

The obtained NB4.05 and $M'$ fluxes enable us to extend the SED of PDS~70~b into the 4--5~$\mu$m regime. To construct the SED, we adopted the $Y$ to $H$ band spectrum from \citet{mueller2018}, which had been obtained with the integral field spectrograph (IFS) of SPHERE \citep{claudi2008}, and also the NIRC2 $L'$ photometry from \citet{wang2020}. These data were combined with the new NB4.05 and $M'$ fluxes from this work, and the reanalyzed photometry of the NACO $L'$ and the SPHERE/IRDIS $H23$ and $K12$ data. For consistency in the SED analysis, we recalibrated the NIRC2 $L'$ magnitude with the stellar spectrum from Sect.~\ref{sec:data_processing} to $14.59 \pm 0.18$. In the $K$ band, we only considered the reanalyzed SPHERE photometry while the SINFONI spectrum from \citet{christiaens2019} was excluded due to a discrepancy in the calibrated fluxes between these datasets (see top panel in Fig.~\ref{fig:sed}). Finally, we adopted a root mean square (rms) noise at 855~$\mu$m of 18~$\mu$Jy~beam$^{-1}$ (i.e., $7.4 \times 10^{-23}$~\flux) from the ALMA continuum imagery by \citet{isella2019} as the approximate ``forced photometry'' \citep[see discussion by][]{samland2017} at the position of PDS~70~b.

The SED is shown in the top panel of Fig~\ref{fig:sed} across the 1--5~$\mu$m range. Apart from the potential broad, H$_2$O absorption feature around 1.4~$\mu$m \citep{mueller2018}, we could not identify any obvious other molecular features (e.g., H$_2$O, CH$_4$, or CO) in the SED on visual inspection. Such absorption features might to be expected given the constraints on the temperature of the atmosphere, which is comparable to the HR~8799 planets \citep[cf.][]{bonnefoy2016,greenbaum2018,molliere2020}. We note that some of the smaller fluctuations in the SPHERE and SINFONI spectra may possibly be attributed to correlated noise. With this in mind, we attempt a simplified fitting approach by describing the spectrum with one or two blackbody components \citep[see also][]{wang2020}. A spectrum based on a single blackbody temperature may naturally describe a photosphere in which the dust opacity dominates over line absorption, with the temperature set either by the internal luminosity of the planet or by the accretion luminosity (see discussion in Sect.~\ref{sec:planet_vicinity}). Later on, a second temperature component is included to account for excess emission at thermal wavelengths ($\gtrsim$3~$\mu$m), for example due to to reprocessed radiation in a CPD.

The fit of the photometric and spectroscopic data was done with \texttt{species}. The toolkit uses the nested sampling implementation of \texttt{MultiNest} \citep{feroz2008} through the Python interface of \texttt{PyMultiNest} \citep{buchner2014}. For the parameter estimation, we used a Gaussian log-likelihood function \citep[see][]{greco2016},
\begin{equation}
\label{eq:log_like}
    \ln \mathcal{L} (D|M) = -\frac{1}{2}\left[(\mathbf{S_\mathrm{IFS}}-\mathbf{F})^{T} \mathbf{C}^{-1} (\mathbf{S_\mathrm{IFS}}-\mathbf{F}) + \sum_{i=1}^{9} \frac{(d_i - m_i)^2}{\sigma_i^2}\right],
\end{equation}
where $D$ is the data, $M$ the model, $\mathbf{S_\mathrm{IFS}}$ the IFS spectrum, $\mathbf{F}$ the model spectrum, $\mathbf{C}$ the (modeled) covariances for the IFS spectrum (see Eq.~\ref{eq:covariances}), $d_i$ the photometric flux for filter $i$, $m_i$ the synthetic flux from the blackbody spectrum, and $\sigma_i$ the uncertainty on the flux $d_i$. The second term of Eq.~\ref{eq:log_like} contains the sum over the nine photometric fluxes that were included in fit.

Spectra from integral field units are known to be affected by correlated noise \citep{greco2016}. We therefore follow the approach by \citet{wang2020} and model the covariances of the SPHERE spectrum as a Gaussian process with a squared exponential kernel \citep{czekala2015,wang2020},
\begin{equation}
\label{eq:covariances}
    C_{ij} = f^2 \sigma_i \sigma_j \exp{\left(-\frac{(\lambda_i - \lambda_j)^2}{2\ell^2}\right)} + (1 - f^2) \sigma_i \sigma_j \delta_{ij},
\end{equation}
where $C_{ij}$ is the covariance between wavelengths $\lambda_i$ and $\lambda_j$, $\sigma_i$ the total uncertainty on the flux of wavelength $\lambda_i$, $f$ the relative amplitude of correlated noise with respect to the total uncertainty, and $\ell$ the correlation length. The correlation length and amplitude were fitted while adjusting the covariance matrix in the log-likelihood function (see Eq.~\ref{eq:log_like}). Finally, each model spectrum was smoothed with a Gaussian filter to match the spectral resolution of the IFS data ($R = 30$) and resampled to the IFS wavelengths with \texttt{SpectRes} \citep{carnall2017}.

\subsubsection{Parameter estimation and model evidence}
\label{sec:parameter_estimation}

The posterior distributions of the temperature, radius, and calibration parameters were sampled with 5000 live points and using uniform priors for all parameters except the correlation length. For the latter, we used a log-uniform sampling of the prior space. The marginalized distributions are shown in Figs.~\ref{fig:posterior_1} and \ref{fig:posterior_2} of Appendix~\ref{sec:posterior} for the cases of fitting one and two blackbody components, respectively. A comparison of the best-fit solution, randomly drawn spectra from the posterior, and the data are shown in Fig.~\ref{fig:sed}.

When fitting one blackbody component, we constrained the temperature and radius of the photospheric region to $1193 \pm 20$~K and $3.0 \pm 0.2$~$\RJ$, and we derived from this a luminosity of $\log(L/L\odot) = -3.79 \pm 0.02$. The overall spectral morphology appears well described by blackbody emission except for the deviation between the $J$ and $H$ bands. Also the 3--5~$\mu$m fluxes match reasonably well with the blackbody emission, thereby confirming the findings by \citet{wang2020}. Specifically, the NB4.05 and $M'$ fluxes deviate from the best-fit spectrum by $1\sigma$. For the covariance model that describes the correlated noise in the IFS spectrum, we determined a length scale of $\approx$0.04~$\mu$m and a fractional amplitude of $0.54 \pm 0.19$. While the upper limit on the ALMA band~7 flux was included in the fit, its impact on the retrieved parameters is negligible because all the single-blackbody model spectra are below the rms noise at band 7.

Although the $M'$ flux only deviates by $1\sigma$ from the best-fit model, we also attempted a fit with two blackbody components to test if such a spectrum provides a better match at wavelengths $\gtrsim$4~$\mu$m. A second blackbody component could for example describe the excess emission from a CPD, which will be discussed in Sect.~\ref{sec:cpd}. Here, we restricted the temperature and radius of the second component to values that are smaller and larger, respectively, than the first component by rejecting samples that did not met this condition. We also restricted the temperature prior of the second component to 0--600~K and the radius to 1--350~\RJ, that is, extending up to $\sim$0.1 times the Hill radius for a 1~\MJ planet at 22~au \citep{tanigawa2012}.

When fitting two blackbody components, the retrieved temperature ($T_1 = 1194 \pm 20$~K) and radius ($R_1 = 3.0 \pm 0.2~\RJ$) of the first component are very similar to those from fitting a single blackbody component. For the second component, we constrained the temperature to $T_2\lesssim256$~K and the radius to $R_2\lesssim245~\RJ$. The sparse wavelength coverage and large uncertainties at wavelengths longer than 4~$\mu$m leave a degeneracy between the temperature and radius of the second component (see Fig.~\ref{fig:posterior_2}). Specifically, a large fraction of the samples is only driven by the upper limit at 855~$\mu$m while not fitting the $M'$ flux, since it is only a $1\sigma$ deviation from the first blackbody component. The posterior of $T_2$ peaks toward 0~K, which is fully degenerate with the radius, $R_2$, going to large values. Therefore, in the bottom panel of Fig.~\ref{fig:sed}, we selected random samples with $T_2>100$~K since the surface layers of a CPD are expected to be heated by accretion \citep[e.g.,][]{aoyama2018}. When considering all posterior samples, we derived a luminosity ratio of $\log(L_1/L_2) = 0.7^{+1.8}_{-1.0}$ for the two components (see Fig.~\ref{fig:posterior_2}). Thus the luminosity of the second component would be about an order of magnitude smaller than the first component.

In addition to the parameter estimation, nested sampling has the advantage of providing the marginalized likelihood (i.e., the model evidence), which enables pair-wise model comparisons. The Bayes factor is used to quantify the evidence of favoring a certain model, and is given by the ratio of the evidence of two models in case the prior probability is the same for both models,
\begin{equation}
\label{eq:bayes_factor}
    B = \frac{\mathcal{Z}(D|M_0)}{\mathcal{Z}(D|M_1)},
\end{equation}
where $\mathcal{Z}(D|M_i)$ is the evidence of data $D$ given model $M_i$. In our case, the Bayes factor is calculated from the evidence ratio of fitting the SED with one or two blackbody components. We obtained a Bayes factor of 2.3, which indicates weak evidence for favoring a model with one blackbody component when considering the Jeffreys' scale \citep[e.g.,][]{trotta2008}.

\section{Discussion}
\label{sec:discussion}

\begin{figure*}
\centering
\includegraphics[width=\linewidth]{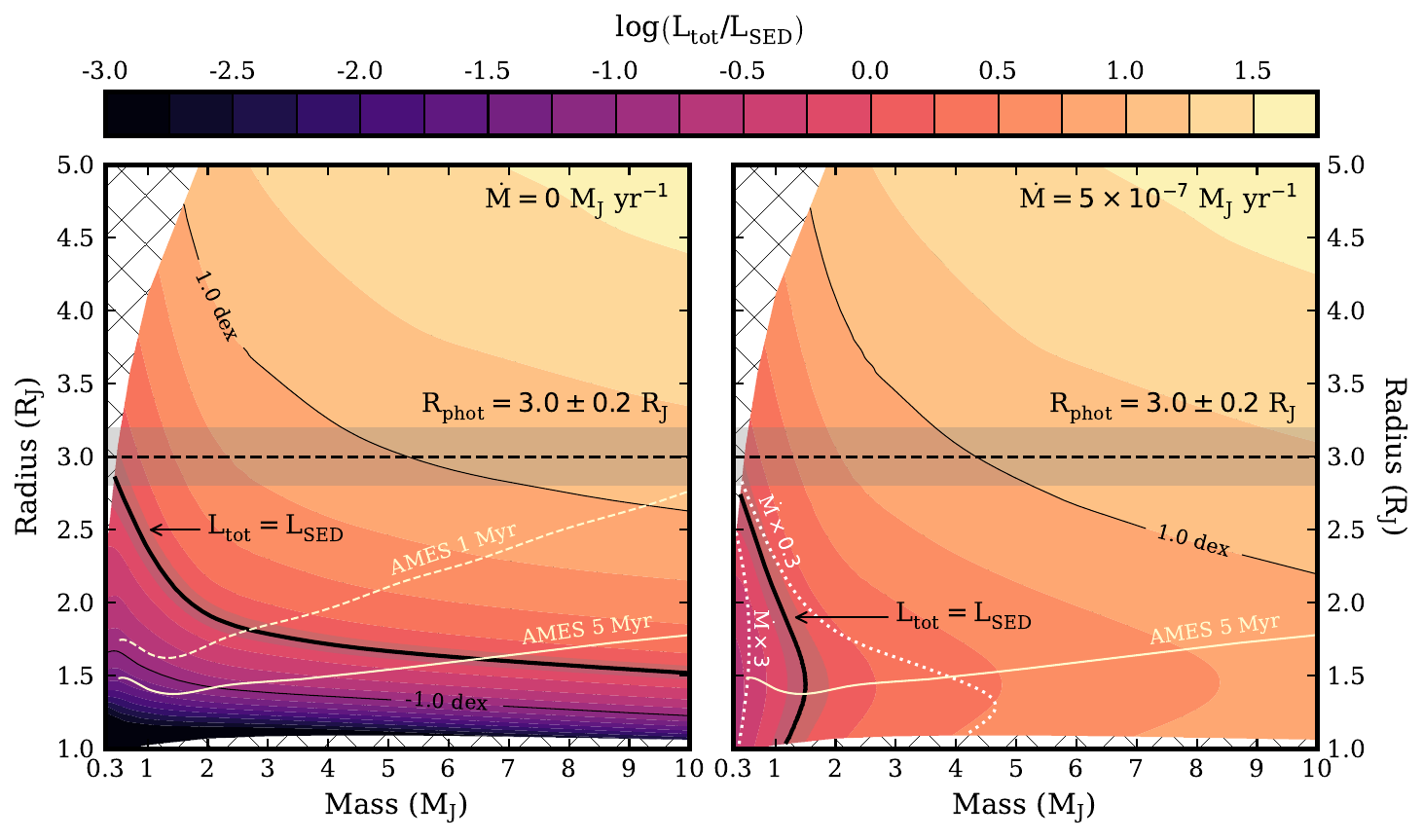}
\caption{Total luminosity from standard models of isolated planets as a function of mass and radius, but with also the accretion luminosity from the planet surface shock (Eq.~\ref{eq:luminosity} with $\LCPD=0$). The \emph{thick black contour} and \emph{gray shade} highlight the measured luminosity and $5\sigma$ uncertainty (see main text for details), $\log(\LSED/\Lsun)=-3.79 \pm 0.02$, and the \emph{thin black contours} are steps of 1.0~dex relative to $\LSED$. The accretion rate was set to $\Mdot=0$ in the \emph{left panel} and $\Mdot=5\times10^{-7}~\MJyr$ \citep{hashimoto2020} in the \emph{right panel}. The inferred photospheric radius and $1\sigma$ uncertainty, $\Rphot\pm\sigma$, are shown with a horizontally \emph{dashed line} and \emph{gray shaded area}. The \emph{light yellow curves} show as reference the mass--radius predictions by the AMES (Cond/Dusty) structure model at 5~Myr (\emph{solid}) and 1~Myr (\emph{dashed}). The \emph{dotted white lines} in the \emph{right panel} show $\Ltot = \LSED$ with $\Mdot$ scaled by a factor of 0.3 and 3. The \emph{crosshatched regions} indicate parts of the parameter space for which no predictions are available from the structure model: because of electron degeneracy pressure at small radii, and because no stable hydrostatic structure exists at large radii.}
\label{fig:luminosity}
\end{figure*}

\subsection{Implications from the luminosity and photospheric radius}
\label{sec:sed_constraints}

Summarizing the fits to one or two blackbody component(s), the blackbody emission radius of the component peaking at smaller wavelength (i.e., $L$ in Fig.~\ref{fig:posterior_1} and $L_1$ in Fig.~\ref{fig:posterior_2}) is $\Rphot\approx3.0\pm0.2~\RJ$, while the corresponding luminosity is $\log(\LSED/\Lsun) = -3.79 \pm 0.02$. Both numbers are comparable to the results of \citet{wang2020} for one blackbody, two blackbodies (taking the luminosity only of the first), and even, as an extreme, their fit to the BT-Settl models. Thus, our \Rphot and \LSED seem robust. Here, we want to analyze systematically what they imply for the physical properties of PDS~70~b.

Since PDS~70~b is presumably still forming, one should think carefully about the evolutionary track models used for the analysis. An important aspect is the time evolution of the models. Cooling tracks need to assume an initial entropy for a given mass and thus, equivalently, an initial radius and luminosity \citep[e.g.,][]{arras2006,marleau2014}. By definition, this 
state of the planet is ``initial'' with respect to the phase of pure cooling. It is set by the formation process and thus also referred to as the ``post-formation'' state \citep{marleau2014}. As pointed out by \citet{fortney2005} and \cite{marley2007} and discussed for instance by \citet[][their Sect.~8.1]{mordasini2012_I}, different formation scenarios will lead to different post-formation entropies. Therefore, the time label used in cooling track models is not guaranteed to be meaningful at early times. This holds in particular for a planet in the middle of formation, as PDS~70~b could conceivably be, but also if the current accretion rate is negligible such that the planet is evolving at essentially constant mass.

Predictions for the entropy of forming and ``newborn'' planets do exist \citep{mordasini2017} but here we take a more general approach. We ignore any time information and consider a grid of hydrostatic gas giant models labeled only by mass, $\Mp$, and radius, $\Rp$. This is possible because for a given atmospheric model, a non-irradiated gas giant planet has only two independent parameters, as discussed in \citet{arras2006}, and also \citet{marleau2014}. In that latter work, these were $\Mp$ and the entropy, $s$, with $\Rp$ or the luminosity seen as functions of $\Mp$ and $s$, while here we consider $\Mp$ and $\Rp$ to be the two independent parameters. This allows us to drop the time label, thus circumventing the uncertainties about \mbox{cold-,} \mbox{warm-,} or hot-start conditions that are linked to the relevant physical processes \citep[e.g.,][]{mordasini2013,berardo2017a,marleau2017,marleau2019b}. Therefore, our approach is independent of the entropy during and at the end of formation.

For the interpretation of the results, we assume that the observable bolometric luminosity of the one blackbody or both is in general the sum of three components:
\begin{equation}
\label{eq:luminosity}
  \Ltot = \Lint(\Mp,\Rp) + \Lacc(\Mp,\Rp,\Mdot) + \LCPD,
\end{equation}
where \Lint is the luminosity from the planet's interior, possibly including some compression luminosity below the surface in the case that there is an accretion shock \citep{berardo2017a}; $\Lacc = \eta G\Mp\Mdot\left(1/\Rp-1/\Racc\right)$ is the luminosity from accretion at the surface of the planet; and \LCPD is the sum of any thermal \citep[e.g.,][]{zhu2015,eisner2015} and shock \citep[e.g.,][]{aoyama2018} emission from a possible CPD. This assumes that any shock luminosity from the planet surface or CPD is reprocessed and thermalized, with only a negligible fraction escaping at least as H$\alpha$; \citet{aoyama2018} report that for a planetary shock, only a small fraction of \Lacc goes into H$\alpha$. In the classical, highly simplified picture of material going directly from the CSD to the planet, the accretion radius is $\Racc\sim\RHill$ \citep{bodenheimer2000}, so that the $1/\Racc$ term is negligible compared to $1/\Rp$. If however the gas releases part of its potential energy between the CSD and the CPD, the effective \Racc would be closer to \Rp but still possibly somewhat larger. Finally, we assume complete local radiative efficiency at the shock, $\eta\approx1$, following the results of \citet{marleau2017,marleau2019b}.

Therefore, in the following, we analyze what requiring $\Ltot=\LSED$ implies. Here, we explore the case in which there is no CPD present or the emission from the CPD is negligible in the total luminosity budget, that is, $\LCPD=0$, motivated by the lack of evidence for a second blackbody component in the SED (see Sect.~\ref{sec:parameter_estimation}). We also assume that $\Racc\gg\Rp$. Alternative scenarios in which $\LCPD$ contributes to the bolometric luminosity will be discussed in Sect.~\ref{sec:cpd}. Finally, we note that Eq.~(\ref{eq:luminosity}) is valid under the assumption of isotropic radiation while, in particular for the shock (\Lacc), this may not be accurate. We will deal with this in a crude manner below by considering also the case $\Lacc=0$.

We will use the BEX-Cond models (Bern EXoplanet cooling tracks; \citealp{marleau2019a}), which graft the AMES-Cond atmospheres \citep{allard2001,baraffe2003} onto the standard Bern planet structure code \texttt{completo21} \citep{mordasini2012_I,mordasini2012_II,linder2019}. The precise choice of atmospheric model, such as AMES-Cond, AMES-Dusty, or that of \citealp{burrows1997}, does not influence much the mapping from \Mp and \Rp to \Lint. In fact, AMES-Cond and AMES-Dusty both use exactly the same $\Rp(t)$ and $\Lint(t)$ tracks (these models differ only in the photometric fluxes), and apart for systematic shifts the results would be very similar for another set of atmospheres. The most important and generic feature of such models is that \Lint increases with both \Mp and \Rp. In general, the functional form of this dependency $\Lint(\Mp,\Rp)$ is different than that of $\Lacc(\Mp,\Rp) \propto \Mp/\Rp$.

\subsubsection{Constraints from the luminosity on the planetary radius and mass}
\label{sec:radius_mass}

We explore first what only the derived luminosity \LSED implies for the physical radius of PDS~70~b, defined as the (very nearly) hydrostatic structure terminating in general at the photosphere or at the shock location. We will return to \Rphot only in Sect.~\ref{sec:radius_comparison} and~\ref{sec:planet_vicinity}. As a limiting case, we consider at first $\Lacc=0$ in Eq.~(\ref{eq:luminosity}), that is, we assume that by some geometric effects (for instance magnetospheric accretion at the planet's poles, away from the observer) most of the (reprocessed) accretion luminosity is not reaching an observer on Earth, and therefore that the observed luminosity is coming only from the photosphere while \LCPD is assumed to be zero. The left panel of Fig.~\ref{fig:luminosity} shows the \Mp and \Rp combinations consistent with this extreme assumption, that is, the models that have $\Ltot=\Lint$ equal to \LSED. At any mass there is an \Rp such that $\Ltot=\LSED$, with $\Rp\approx2.8~\RJ$ at most, down to 1.5~\RJ at 10~\MJ. Formally, there is no upper mass limit.

The right panel of Fig.~\ref{fig:luminosity} assumes instead $\Mdot=\Mdotmin=5\times10^{-7}~\MJyr$ in Eq.~(\ref{eq:luminosity}), corresponding to the lower limit on the accretion rate derived by \citet{hashimoto2020}. Now, only small masses $\Mp\lesssim1.5~\MJ$ are allowed since \Lacc raises everywhere \Ltot significantly. For example, at $\Mp=5~\MJ$, \LSED would need to be higher than derived by at least $\approx$0.5~dex ($\approx$25$\sigma_{\LSED}$) for there to be a matching luminosity. The discrepancy is larger for larger \Mp values. Where \Ltot can be matched, however, the possible radii\footnote{For a narrow mass around $\Mp\approx1$--$1.5~\MJ$, there are two solutions: a small- and a large-\Rp solution, with, respectively, a small (large) \Lint and large (small) \Lacc, summing up to $\Ltot=\LSED$.} range from $\Rp\approx1.1$ to~$\approx2.8~\RJ$. This is thus the physical radius of PDS~70~b implied by \LSED alone (i.e., ignoring \Rphot), assuming $\Mdot=5\times10^{-7}~\MJyr$, and the corresponding mass is $\Mp\approx0.5$--1.5~$\MJ$.

As a rough check, we inspected the Bern population synthesis\footnote{The data can be visualized at and downloaded from the Data Analysis Centre for Exoplanets (DACE) platform at \url{https://dace.unige.ch}.}, both from Generation~Ib \citep{mordasini2012_I,mordasini2012_II,mordasini2017} and from the newest, Generation~III \citep{emsenhuber2020a,emsenhuber2020b} to see whether this combination of $(\Mdot,\Mp,\Rp)$ is met. We find that, not considering the time at which this happens in the population synthesis, planets accreting at $\Mdot\approx\Mdotmin$ have $\Rp\approx1.3$--$1.7~\RJ$ for $\Mp\approx0.5$--$2~\MJ$, reaching up to $\Rp\approx2.5~\RJ$ down to $\Mp\approx0.3~\MJ$. Since this range of \Rp is within our allowed range $\Rp\approx1.1$--$2.8~\RJ$, the \Mdot would be consistent with these formation models.

As mentioned, the \Mdot value from \citet{hashimoto2020} is a lower limit. Already at $\Mdot\approx3\Mdotmin$, the implied mass from the structure models (dotted white line in the right panel of Fig.~\ref{fig:luminosity}) is\footnote{At these low masses, the structure models become sensitive to other parameters such as metallicity or core mass, so that this value is to be taken with a grain of metal. However, that no high-mass models are possible here is robust.} $\Mp\lesssim0.6~\MJ$. This mass might seem small but at least in the Bern population synthesis, there are planets in the corresponding region of $(\Mdot,\Mp,\Rp)$, again not taking time into account. Thus such low-mass solutions might be possible. On the other hand, if the lower limit on \Mdot is overestimated, then the derived \Mp is underestimated (see the $\Mdot = 0.3\Mdotmin$ case in Fig.~\ref{fig:luminosity}) because $\Ltot \approx \Lacc \propto \Mp\Mdot$. Hence, to keep the luminosity constant (i.e., equal to \LSED), a smaller \Mdot is compensated by an increase in \Mp. Nevertheless, we need to see whether the derived \Mp range matches other constraints.

One constraint is the presence of a gap. From \citet{kanagawa2016} a suitable combination of disk parameters (scale height and viscosity) could lead to a gap even at low masses. For example, with an aspect ratio of $H_\mathrm{p}/\rp = 0.067$ \citep{bae2019} at the separation of the planet ($\rp\approx22$~au), an estimated gap width of 20~au, and a turbulence parameter of $\alpha = 10^{-4}$, given the evidence for weak turbulence in protoplanetary disks \citep[e.g.,][]{flaherty2020}, the relation from \citet{kanagawa2016} implies a planet mass $\Mp\approx1~\MJ$. For this combination of parameters, the derived mass is likely an upper limit since the gap in the PDS~70 disk is opened by the combined effect of two planets. In any case, in a first approximation the low \Mp value inferred from the luminosity seems compatible with the presence of a gap. More detailed, radiation-hydrodynamical modeling of the disk would clearly be warranted. 

A second aspect concerns the orbital stability of the system. \citet{bae2019} studied the dynamics of the PDS~70 system by fixing $\Mp=5~\MJ$ for planet~b and varying the mass of~c from 2.5~to 10~$\MJ$. They concluded that the orbits are likely in a 2:1 mean-motion resonance (as had been suggested by \citealt{haffert2019}) and can remain dynamically stable over millions of years. It would be interesting to repeat their simulations with a lower mass $\Mp\approx1~\MJ$ for planet~b, possibly also considering a lower mass for PDS~70~c. A low \Mp value would be in agreement with the $N$-body simulations by \citet{mesa2019}, who showed that the two-planet system would be stable with masses of 2~\MJ, whereas dynamical perturbations occurred in their simulations with higher-mass planets.

Coming back to the luminosity constraint, we compare \LSED to the AMES (hot-start) isochrones, discussing the validity of this approach afterward. Figure~\ref{fig:luminosity} shows that \LSED intersects the 5-Myr isochrone (roughly the age of the star) at $\Rp=1.6~\RJ$ when not considering a contribution of \Lacc in \Ltot (left panel), and at $\Rp\approx1.4$--1.5~\RJ for $\Mdot\approx(0.3$--$3)\Mdotmin$ (right panel). Taken at face value, the corresponding masses are $\Mp\approx6.5~\MJ$ for $\Lacc=0$ and again $\Mp\approx0.5$--1.5~$\MJ$ for $\Mdot\approx(3$--$1)\Mdotmin$.

However, several caveats apply. One is that the AMES models were made for isolated planets, whereas during formation there can be a spread of at the very least 0.3~dex in \Lint at a given mass at 5~Myr for what are effectively hot starts (see Figs.~2 and~4 of \citealt{mordasini2017}). This will affect the derived radius. Another concern is that there might be a formation delay of perhaps a few~Myr, which would be significant at this age \citep{fortney2005}. In the case of HIP~65426, which has a mass of $2~\Msun$ \citep{chauvin2017}, \citet{marleau2019a} estimated roughly a formation time near 2~Myr and argued that this should increase with lower stellar mass (PDS~70 has a mass of 0.8~$\Msun$; \citealt{keppler2018}). Finally, the true shape of the physical isochrone could be different than in the AMES track, which was not guided by a formation model. In particular the post-formation radius as a function of mass could conceivably be non-monotonic, allowing for several solutions to $\Ltot(t=5~\mathrm{Myr})=\LSED$.

We show as an extreme comparison the 1-Myr hot-start isochrone in the left panel of Fig.~\ref{fig:luminosity}. This would imply a somewhat larger radius $\Rp\approx1.8~\RJ$. For $\Lacc=0$, the mass would be clearly smaller, with $\Mp\approx3~\MJ$, whereas for $\Mdot=\Mdotmin$ the mass would be similar to the 5~Myr isochrone (not shown explicitly in the plot). We note that the AMES isochrone at the maximal age (the system age) does provide an upper mass limit within the hot-start assumption; younger ages necessarily imply smaller masses, as Fig.~\ref{fig:luminosity} makes clear. However, the initial radius could be smaller. While extreme cold starts \`a la \citet{marley2007} are disfavored \citep[e.g.,][]{berardo2017a,mordasini2017,snellen2018,wang2018,marleau2019b}, warm starts seem a realistic possibility. In this case, a 5-Myr isochrone would match the luminosity at a smaller radius and \emph{larger} mass---thus the mass upper limit from the hot start is in fact a ``lower upper limit,'' meaning it is not informative.

Interestingly, the mass that we derived from the luminosity \LSED and the adopted lower limit on the accretion rate (i.e., the right panel in Fig.~\ref{fig:luminosity}) appears lower then what has been inferred in previous studies. \citet{keppler2018} compared the $H$ and $L'$ band magnitudes with predictions by evolutionary models and estimated a mass of 5--9~\MJ and 12--14~\MJ with a hot-start and warm-start formation, respectively. Similarly, the estimates by \citet{mueller2018} also assumed that the NIR fluxes trace directly the planet atmosphere. The authors determined a mass of 2--17~\MJ by fitting the SED with atmospheric model spectra. One difference in our analysis is that it is based on the bolometric luminosity, but the key point is that it takes the accretion luminosity into account. Therefore, it is not surprising that we derive a different (i.e., lower) planet mass. We note that our quoted mass error bars are smaller both in relative and absolute terms with respect to previous studies, but our error bars do not reflect the (large) uncertainty on the accretion rate.

More recently, \citet{wang2020} estimated a mass for PDS~70~b of 2--4~\MJ from the bolometric luminosity by using the evolutionary models of \citet{ginzburg2019} and assuming a system age of 5.4~Myr. Our mass constraint ($\Mp \approx0.5$--1.5~$\MJ$ with $\Mdot=\Mdotmin$) is somewhat comparable with the findings by \citet{wang2020}, indicating a relatively low mass compared to earlier estimates. However, we want to stress again that the low planet mass that we estimated hinges on the adopted accretion rate. If the accretion rate is overestimated then the mass is underestimated, as can be seen from the $\Mdot \times 0.3$ example in Fig.~\ref{fig:luminosity}. The mass of planet b that was derived by \citet{hashimoto2020} from the H$\alpha$ flux is significantly larger ($\sim$12~\MJ) than our value. However, the authors noted that the line profile was not resolved, hence only an upper limit on the free-fall velocity could be determined ($v_0 = 144$~km~s$^{-1}$). Since the planet mass scales quadratically with the velocity (see Eq.~3 in \citealt{hashimoto2020}), it would require a factor of $\approx$3 smaller velocity to lower the estimated mass from 12 to 1.5~\MJ. We note that a velocity of $v_0 = 48$~km~s$^{-1}$ would still be twice as large as the minimum velocity that is required to produce H$\alpha$ emission (i.e., $v_{0,\mathrm{min}}\approx 25$~km~s$^{-1}$; see Fig.~6 by \citealt{aoyama2018}).

\subsubsection{Comparing the planetary and photospheric radii}
\label{sec:radius_comparison}

How does the planet radius discussed so far compare to the derived photospheric radius derived above, $\Rphot\approx3.0~\RJ$? In both panels of Fig.~\ref{fig:luminosity}, and in particular for $\Mdot\gtrsim\Mdotmin$, the models with $\Ltot=\LSED$ all have $\Rp < \Rphot$, with a substantial difference between the two. Put differently, there is no mass predicted by the structure model for which the radius is equal to $\Rphot$ \emph{and} the total luminosity is equal to $\LSED$ simultaneously. A non-zero contribution from an accretion luminosity only exacerbates the tension.

For a range of masses between 1~and $10~\MJ$, the discrepancy $\DR\equiv\Rphot-\Rp$ is typically at least 0.7--$1.5~\RJ$, which is $3.5\sigma_{\Rphot}$ at $\Mp=1~\MJ$ and $7.5\sigma_{\Rphot}$ at $\Mp=10~\MJ$. It does decrease toward low masses for both $\Lacc=0$ and $\neq0$, such that formally there is a narrow match within the 1--2$\sigma$ regions of \Rphot and \LSED. However, this is at the maximum radius possible for a convective hydrostatic planet and thus seems unlikely, especially given that PDS~70~b probably has evolved at least for a short time, even if not the full age of the system (5.4~Myr). In short, there is in fact no satisfying solution within one or two~$\sigma_{\Rphot}$.

If one takes the AMES isochrone at 5~Myr, the discrepancy between the physical and photospheric radii is at the very least (taking $\Lacc=0$) $\Delta R=1.4~\RJ$, or $\approx7\sigma_{\Rphot}$. With the extreme case of a 4.4-Myr formation delay, and thus the 1-Myr isochrone, the difference is still $6\sigma_{\Rphot}$. In any case, we argued that the AMES cooling models are possibly not directly appropriate for (maybe still forming) young planets.

There are four non-mutually exclusive possible implications from this discrepancy between the inferred physical and photospheric radii:
\begin{enumerate}
    \item[(i)]   \LSED is underestimated. This could be the case if \Lacc dominates \LSED (after reprocessing) and is emitted anisotropically, which is conceivable. A dominating \Lacc in turn seems plausible if \Mdot is higher than \Mdotmin from \citet{hashimoto2020}. Alternatively, or in addition, extinction in the system may lead not only to radiation (\Lint and/or \Lacc) being shifted to longer wavelengths, but also to it being \emph{re}-emitted away from the observer. This could possibly come from non-isotropic scattering by dust grains in the CSD (through the upper layers of which we are observing the PDS~70~b region) or in a CPD. In any case, such geometry effects would let \LSED represent only a fraction of \Ltot, allowing in principle for mass--radius solutions given classical planet structure models.
    
    \item[(ii)]  \Rphot is overestimated. Assuming that the data constrain the shape and thus the approximate \Teff of the spectrum, this is equivalent to~(i). The derived \Teff and \Rphot from fitting synthetic spectra are sometimes correlated, therefore a different model spectrum may give a larger \Teff and a smaller \Rphot, without changing \LSED much. For example, a decrease in the radius by 50\% would correspond to an increase in the temperature by $\approx$40\%, such that the luminosity remains constant. Alternatively, extinction by small dust grains could have altered the SED, possibly mimicking a larger radius and smaller temperature (see Fig.~\ref{fig:color_magnitude}).

    \item[(iii)] The structure models (classical, non-accreting gas giants that turn out to be fully convective) do not apply and \Lint should be smaller at a given $(\Mp,\Rp)$, or equivalently \Rp should be larger at a given $(\Mp,\Lint)$, assuming \Lint still increases with both \Mp and \Rp. Recent modeling work by \citet{berardo2017a} suggests that this might hold, at least qualitatively, but the effect might not be large enough.

    \item[(iv)]  \Rp \emph{is} the physical radius of the planet, implying there is Rosseland-mean optically thick material between \Rp and \Rphot.
\end{enumerate}
The last possibility is a particularly interesting one that we consider in more detail in the next subsection.
 
\subsubsection{Constraints on the vicinity of the planet?}
\label{sec:planet_vicinity}

We will assume that there is optically thick material between \Rp and \Rphot. This material could be flowing onto the planet or be in some layer of the CPD. However, given that CPDs are thought to be a fraction of the size of the Hill sphere \citep[e.g.,][]{lubow2009} and that $\Rphot\ll\RHill\sim3500~\RJ$ for a $\sim$1~\MJ planet at 22~au, this is most likely material flowing onto the planet (see Sect.~\ref{sec:cpd} for a more detailed discussion on the presence/absence of a CPD).

In principle, the extinction could be due to gas or to dust opacity. However, the absence of strong molecular features (as argued in Sect.~\ref{sec:modelling_approach}), contrary to what would be expected from gas at $T\sim1000$~K, suggests that the opacity is grayer and dust-dominated \citep{wang2018}. One can estimate whether the (possibly high) temperatures near the planet would allow for dust to exist within $\Rphot\approx3.0~\RJ$. From \citet{isella2005}, dust is destroyed at $\Tdest\approx1280$--1340~K at $\rho\sim10^{-10}$--$10^{-9}$~g\,cm$^{-3}$ (see below). Assuming that the luminosity is approximately constant in the accretion flow onto the planet \citep{marleau2017,marleau2019b}, $\Teff=1193$~K at $\Rphot=3.0~\RJ$ implies a local $\TeffL=1307$~K at a radius $2.5~\RJ$ ($\Teff=1687$~K at $1.5~\RJ$). The temperature of the gas and dust, in turn, is given by solving the implicit approximate equation $T\approx\TeffL/4^{1/4}\times(1+1.5\kapR\rho\Rp)^{1/4}$, where $\kapR(T)$ is the Rosseland mean opacity (see Eq.~(32) of \citealp{marleau2019b}). Given the numerical values, we estimate that at both positions the dust could be partially destroyed. This suggests that the dust destruction, which is strongly sensitive to the temperature \citep[][]{bell1994,semenov2003}, could occur over a non-negligible spatial scale comparable to $\Delta R$. This effect is seen as the temperature plateau in Fig.~9 of \citet{marleau2019b}. Geometrical effects in the accretion flow will affect the details but in an average sense, the region between the planet radius and the photosphere could well be partially filled with dusty material, with full abundance near \Rphot decreasing smoothly toward \Rp.

For the radiation to be reprocessed between \Rp and \Rphot, there must be at least a few Rosseland-mean optical depths between the two radii, that is, $\Delta\tauR\sim\langle\kapR\rho\rangle (\Rphot-\Rp) > 1$ (i.e., the infrared extinction must be $A_\mathrm{IR} \gtrsim 1$~mag). For a filling factor \ff, the typical gas preshock density is $\rho=\Mdot/(4\pi\Rp^2\ff\vff)\sim10^{-10}$~g\,cm$^{-3}$, with the preshock free-fall velocity given by $\vff^2\approx 2G\Mp/\Rp$ \citep[e.g.,][]{zhu2015}. With $\Delta R\sim1~\RJ$,
the requirement $\Delta\tauR>1$ implies $\kapR\gtrsim1$~\opacity as a conservative lower limit. This is the opacity per unit gas mass, that is, the opacity per unit dust mass \kapRm times the dust-to-gas ratio \fpg. At these temperatures near or below 2000~K, the gas opacity is $\kapR\lesssim10^{-2}$~\opacity \citep{malygin2014}, which implies that dust dominates the total opacity budget, so that the requirement would be $\kapRm\gtrsim100/(\fpg/0.01)$~\opacity. For the canonical $\fpg=0.01$, this needed opacity is in line with the calculation of \citet{semenov2003} and seems in general reasonable given the uncertainties about the exact dust composition, porosity, non-sphericity, material properties, etc. If the accreting gas comes from high latitudes in the CSD \citep{tanigawa2012,morbidelli2014,teague2019}, the settling of the dust to the midplane could imply a lower \fpg in the accretion flow \citep{uyama2017}, perhaps by as much as a few orders of magnitude. Even in this case, however, the required \kapRm seems consistent with predictions from \citet{woitke2016} for an appropriate size distribution and material properties for the dust. Finally, this rough estimate assumes a spherically symmetric accretion flow; if $\ff<1$ in the accretion flow and this concentration of matter is along the line of sight, the required minimum opacity would be lower, proportionally to $\ff$, and thus easier to reach. Therefore, altogether, a photospheric radius that is larger than the planet radius could be explained by dusty material in the vicinity of PDS~70~b.

\subsection{Mid-infrared excess from a circumplanetary disk?}
\label{sec:cpd}

The formation of a giant planet is characterized by several distinct phases of growth. At an early stage, the accretion flow from the CSD feeds directly the atmosphere of the object, through spherical accretion of gas and solids entering the planet's Hill sphere \citep[e.g.,][]{pollack1996,cimerman2017}. As the planet grows further, the gaseous envelope may collapse, thereby triggering a runaway accretion and the potential formation of a CPD \citep[e.g.,][]{canup2002,dangelo2003a}. Hydrodynamical simulations have indeed shown that a CPD can remain, spanning a fraction of the planet's Hill radius \citep[e.g.,][]{lubow2009,ayliffe2012,tanigawa2012,szulagyi2016}. The disk will act as important mediator for channeling the infalling gas and dust toward the planet \citep[e.g.,][]{tanigawa2012} and the accretion onto the planet--disk system may leave a strong imprint on the bolometric luminosity \citep{papaloizou2005}.

In Sect.~\ref{sec:sed_constraints}, we assumed that the SED luminosity reflected the planet's interior and accretion luminosity, while ignoring a contribution from a CPD. The analysis in Sect.~\ref{sec:sed_analysis} revealed indeed weak evidence that the SED is better described by one blackbody component instead of two. This is mainly because the second component is only constrained by the $1\sigma$ deviation of the $M'$ flux and the non-detection with ALMA at the expected position of PDS~70~b. Therefore, an alternative interpretation based on the deviation of the $M'$ flux will be very speculative. Nonetheless, we will briefly discuss our findings in the context of a CPD that could be present.

If there is no excess emission at MIR wavelengths, the $L'$, NB4.05, and $M'$ fluxes trace the same photospheric region as the NIR part of the SED. In that case, the SED is described by a single temperature and radius, which can be characterized by an extended dusty environment, as discussed in Sect.~\ref{sec:sed_constraints}. A non-detection of a CPD in $M'$ and with ALMA at 855~$\mu$m may indicate that the CPD is either very faint (e.g., low in temperature and/or mass), that the physical conditions near the planet do not allow (yet) the formation of a CPD, or that the CPD may have already been dispersed. This finding, combined with the constraint on the mass of PDS~70~b from Sect.~\ref{sec:radius_mass} ($\sim$1~\MJ) may guide the calibration of CPD models.

Alternatively, we speculate that the slight excess emission in $M'$ could trace a second component from a cooler ($\lesssim$256~K) and more extended region ($\lesssim$245~$\RJ$) that is associated with a CPD. This could either be thermal emission coming directly from the disk or reprocessed emission from the accretion shock on the surface of the disk, as given for example by \citet{aoyama2018}. From the retrieved temperatures and radii of the two blackbody components, we derived that the luminosity of the cooler component, $L_2$, is approximately an order of magnitude smaller than that of the first component, $L_1$ (although they could be comparable within 1$\sigma_{L_1/L_2}$ since $\log(L_1/L_2) = 0.7^{+1.8}_{-1.0}$; see Fig.~\ref{fig:posterior_2}). If we assume that the CPD is heated by the luminosity of the planet, this may indicate that $\lesssim$10\% of the planet flux is reprocessed by the CPD. Here, the percentage of reprocessed emission is an upper limit since the CPD is also expected to be heated by accretion from the CSD and/or viscous heating. Such a process may in fact dominate the luminosity budget of the CPD.

Previously, \citet{christiaens2019} suggested that part of the $K$ band flux originates from a CPD, since the considered atmosphere models could not explain the absolute flux and slope of the SINFONI spectrum. Although there is a discrepancy between SINFONI and SPHERE $K$ band fluxes, Fig.~\ref{fig:sed} shows that the SPHERE photometry is consistent with a blackbody spectrum \citep[see also][]{wang2020}, therefore possibly not requiring excess flux from a CPD at these wavelengths. However, this needs to be confirmed. Instead, we identified very marginal excess emission in the $M'$ band, but we stress that the result is not significant. More precise photometry at 4--5~$\mu$m is required to constrain the circumplanetary characteristics of PDS~70~b, for example with the aperture masking interferometry (AMI) mode of the NIRISS instrument \citep{artigau2014} on board the \emph{James Webb Space Telescope} (\emph{JWST}).

The spectral appearance of a forming planet and the disk surrounding it will deviate from that of an isolated object with atmospheric emission alone. Predictions by \citet{zhu2015}, based on a simplified, steady-state disk model, showed that an accreting CPD can be brighter at near- and mid-IR wavelengths than the planet itself if $\Mp\Mdot$ is sufficiently large. Therefore, the peak in the observed SED may solely trace the hottest region of the CPD instead of the planet atmosphere. This would imply that the actual planet is not visible and we mainly detect the (reprocessed) luminosity from the disk, that is, $\LSED \approx \LCPD$.

Considering a $1~\MJ$ planet and $\Mdot=5\times10^{-7}~\MJyr$ \citep{hashimoto2020}, the predictions in Fig.~1 by \citet{zhu2015} show that the NIR fluxes are dominated by the emission from the planet atmosphere instead of the viscous heating in the CPD (i.e., when $\Mp\Mdot \sim 10^{-7}$--$10^{-6}~\MMJyr$ and $\Teff \sim 1000$~K), unless the inner radius of the CPD is very small ($\Rin = 1~\RJ$) and/or $\Mp\Mdot \gtrsim 10^{-6}~\MMJyr$. In the case the SED mostly traces emission from the CPD, the spectral slope is expected to be less steep at longer wavelengths due to the radial temperature gradient in the disk. This contrasts with the observed SED, which is consistent with a single blackbody (see Sect.~\ref{sec:sed_analysis}), therefore pointing to a photospheric region that is characterized by a single temperature.

Since the adopted accretion rate from \citet{hashimoto2020} is a lower limit, we applied the fitting procedure from Sect.~\ref{sec:modelling_approach} on the predicted magnitudes by \citet{zhu2015} to test what blackbody temperature and radius would be retrieved if the accretion rate is larger and the SED only traces CPD emission. For this, we considered the full-disk case with $\Rin = 2~\RJ$ and $\Mp\Mdot = 10^{-5}~\MMJyr$. We adopted the $J$- to $M'$-band magnitudes and added arbitrary error bars of 0.1~mag. When fitting a single blackbody, we retrieved $\Teff = 994 \pm 15$~K and $R = 11.6 \pm 0.6~\RJ$. The flux density peaks at $\sim$3~$\mu$m, which is similar to the SED of PDS~70~b (see top panel in Fig.~\ref{fig:sed}), and the $M'$ flux from the CPD model shows a 10--20\% excess with respect to the best-fit blackbody spectrum (due to the temperature gradient in the disk). Interestingly, while the temperature is somewhat comparable to the photospheric temperature of PDS~70~b ($\Teff = 1193 \pm 20$~K), the retrieved radius from the predicted CPD fluxes is clearly larger than the photospheric radius of PDS~70~b ($R = 3.0 \pm 0.2$~$\RJ$). This brief assessment may suggest that both the accretion rate and photospheric radius of PDS~70~b are too small to interpret the SED as $\LSED \approx \LCPD$, so the photosphere traces presumably a more compact, dusty environment around the planet instead of a CPD. However, a more detailed analysis would be required to confirm this.

Apart from a luminosity contribution by a viscously heated CPD, the accretion flow and shock (on the planet surface and/or disk) may further alter the energy distribution. For example magnetospheric accretion from the disk onto the planet could also heat the photosphere of the planet, thereby enhancing the flux at shorter wavelengths \citep{zhu2015}. The importance of such accretion processes remain poorly constrained and can additionally be variable and subject to outbursts \citep{lubow2012,brittain2020}.

\section{Summary and conclusions}
\label{sec:conclusions}

We have reported on the first detection of PDS~70~b at 4--5~$\mu$m. We used high-resolution observations with NACO at the VLT to image the forming planet with the NB4.05 (Br$\alpha$) and $M'$ filters. PDS~70~c is tentatively recovered in NB4.05 and the near side of the gap edge of the CSD is detected in scattered light. We have also reanalyzed the photometry of PDS~70~b from archival SPHERE $H23$ and $K12$, and NACO $L'$ imaging data.

The absolute $M'$ flux of PDS~70~b is compatible with a late M-type dwarf, and the young, planetary-mass objects ROXs~42~Bb and GSC~06214~B. The NIR~--~$M'$ colors, on the other hand, are redder than any of the known directly imaged planets and most comparable to the dusty, L-type companions HD~206893~B and HIP~65426~b. While the $M'$ magnitude and related colors are unusual compared to other directly imaged planets, they are consistent with blackbody emission from an extended region that is several times the radius of Jupiter.

With the new NB4.05 and $M'$ photometry, we modeled the available SED data (including a SPHERE/IFS spectrum) by assuming a blackbody and derived a photospheric temperature of $\Teff = 1193 \pm 20$~K and radius of $\Rphot = 3.0 \pm 0.2$~$\RJ$, which is consistent with the blackbody analysis of the 1--4~$\mu$m SED by \citet{wang2020}. Apart from small-scale deviations (partially due to expected correlated noise in the NIR spectra) and the tentative H$_2$O feature at 1.4~$\mu$m, the photometric and spectroscopic data appear to be well described by a single blackbody temperature and radius. From the sampled posterior distributions, we derived a bolometric luminosity of $\log(L/\Lsun) = -3.79 \pm 0.02$.

The derived luminosity and photospheric radius enabled us to place constraints on the planetary radius and mass of PDS~70~b. We used standard models for isolated gas giant planets to infer the mass--radius solutions corresponding to the measured luminosity, while taking into account the accretion luminosity. The time-independent approach of the analysis makes it unaffected by the uncertain cooling time of the object. Here we summarize the main findings and conclusions from this analysis:
\begin{enumerate}
    \item[(i)] In the limiting case that $\Lacc = 0$ (e.g., due to a geometric effect), there are solutions of the radius for all considered masses (up to 10~\MJ), but always smaller than \Rphot.
    \item[(ii)] When including \Lacc in the luminosity budget (based on the \citet{hashimoto2020} estimate of the accretion rate),
    only masses up to 1.5~\MJ have solutions for which the observed luminosity is equal to the combination of the intrinsic and accretion luminosity.
    \item[(iii)] Considering these two cases, we constrain the mass of PDS~70~b to $\Mp\approx0.5$--1.5~\MJ and the physical radius to $\Rp\approx1$--2.5~\RJ. This is consistent with predictions from population synthesis models of forming planets and an approximate estimate based on the gap width.
    \item[(iv)] The discrepancy between the photospheric and planetary radius could imply that the planet is enshrouded by a dusty, extended environment, which is consistent with the approximate blackbody spectrum and the dearth of strong molecular features.
    \item[(v)] The derived photospheric radius is orders of magnitude smaller than the planet's Hill radius. In the case of a dusty envelope, this indicates that the extended region is actively replenished by dust that is coupled to the gaseous accretion flow from the CSD \citep[see also][]{wang2020}.
    \item[(vi)] Alternatively, the discrepancy may indicate that the actual luminosity is larger than the observed luminosity, for example due to anisotropic emission or scattering, extinction, or that the structure models may not apply because PDS~70~b is still forming.
\end{enumerate}

The $M'$ flux shows a slight deviation from the best-fit results when considering a single blackbody temperature. We modeled the MIR excess with a second blackbody component and obtained an approximate upper limit on the temperature and radius of potential emission from a CPD, $\Teff\lesssim256$~K and $R\lesssim245~\RJ$, but the Bayes factor indicates weak evidence that the data is better described by a model with a single blackbody component. Higher-precision photometry at MIR wavelengths is required to place stronger constraints on potential emission from a CPD, for example with the improved 4--5~$\mu$m imaging capabilities of VLT/ERIS, the AMI mode of NIRISS instrument on \emph{JWST}, and in the further future $M'$ and $N$ band photometry with ELT/METIS.

\begin{acknowledgements}

We would like to thank Andr\'e M\"{u}ller, Dino Mesa, and Valentin Christiaens for kindly sharing their SPHERE and SINFONI spectra and we thank Yuhiko Aoyama, Timothy Gebhard, Greta Guidi, and Jason Wang and the ExoGRAVITY team for clarifying discussions. We also thank the referee whose constructive comments improved the quality of this manuscript. T.S. acknowledges the support from the ETH Zurich Postdoctoral Fellowship Program. G.-D.M. acknowledges the support of the DFG priority program SPP~1992 ``Exploring the Diversity of Extrasolar Planets'' (KU~2849/7-1) and from the Swiss National Science Foundation under grant BSSGI0\_155816 ``PlanetsInTime''. T.S., G.C., and S.P.Q. thank the Swiss National Science Foundation for financial support under grant number 200021\_169131. P.M. acknowledges support from the European Research Council under the European Union's Horizon 2020 research and innovation program under grant agreement No. 832428. K.O.T acknowledges support from the European Research Council (ERC) under the European Union’s Horizon 2020 research and innovation programme (grant agreement no. 679633; Exo-Atmos). Part of this work has been carried out within the framework of the National Centre of Competence in Research PlanetS supported by the Swiss National Science Foundation. S.P.Q. acknowledges the financial support of the SNSF.

\end{acknowledgements}

\bibliographystyle{aa}
\bibliography{references}

\begin{appendix}

\section{Astrometric calibration}
\label{sec:astrometry}

In this appendix, we provide an overview of the calibrated astrometry. The final separation is calculated by adding the bias offset and combining the two error components in quadrature. The final position angle is calculated by adding the bias and true north offset and combining the three error components in quadrature. For NACO, we adopted a plate scale of 27.2 mas pixel$^{-1}$ and true north of $-0\ffdeg44 \pm 0\ffdeg1$ from \citet{cheetham2019}. For SPHERE/IRDIS, we adopted from \citet{maire2016b} a plate scale of 12.25 and 12.26~mas~pixel$^{-1}$ for the $H23$ and $K12$ filters, respectively, and true north of $-1\ffdeg75 \pm 0\ffdeg08$.

\begin{sidewaystable}
\caption{Astrometry and error budget.}
\label{table:astrometry}
\centering
\bgroup
\def\arraystretch{1.25}
\begin{tabular}{L{2.5cm} C{2cm} C{2.1cm} C{1.9cm} C{2cm} C{1.8cm} C{2cm} C{2cm}}
\hline\hline
Filter & Separation MCMC & Separation bias & P.A. MCMC & P.A. bias & True north correction & Final separation & Final P.A. \\
 & (mas) & (mas) & (deg) & (deg) & (deg) & (mas) & (deg) \\
\hline
\multicolumn{8}{l}{\emph{PDS 70 b}}\\
SPHERE $H2$ & $173.26 \pm 3.18$ & $0.21 \pm 4.19$ & $154.08 \pm 0.46$ & $0.05 \pm 0.66$ & $-1.75 \pm 0.08$ & $173.47 \pm 5.26$ & $152.37 \pm 0.81$ \\
SPHERE $H3$ & $173.88 \pm 3.16$ & $0.39 \pm 3.69$ & $154.04 \pm 0.42$ & $-0.02 \pm 0.65$ & $-1.75 \pm 0.08$ & $174.28 \pm 4.85$ & $152.27 \pm 0.77$ \\
SPHERE $K1$ & $182.68 \pm 1.14$ & $0.12 \pm 0.95$ & $147.61 \pm 0.17$ & $-0.01 \pm 0.17$ & $-1.75 \pm 0.08$ & $182.80 \pm 1.48$ & $145.86 \pm 0.25$ \\
SPHERE $K2$ & $183.53 \pm 1.68$ & $-0.16 \pm 1.99$ & $146.83 \pm 0.24$ & $0.01 \pm 0.25$ & $-1.75 \pm 0.08$ & $183.37 \pm 2.61$ & $145.09 \pm 0.35$ \\
NACO $L'$ & $207.31 \pm 9.94$ & $-0.22 \pm 10.58$ & $150.29 \pm 1.02$ & $-0.08 \pm 1.32$ & $-0.44 \pm 0.10$ & $207.09 \pm 14.52$ & $149.76 \pm 1.67$ \\
NACO NB4.05 & $205.98 \pm 18.80$ & $2.65 \pm 18.90$ & $148.40 \pm 1.91$ & $-0.07 \pm 2.79$ & $-0.44 \pm 0.10$ & $208.62 \pm 26.66$ & $147.89 \pm 3.38$ \\
NACO $M'$ & $179.70 \pm 11.22$ & $-2.81 \pm 10.22$ & $138.75 \pm 2.39$ & $0.53 \pm 3.83$ & $-0.44 \pm 0.10$ & $176.88 \pm 15.18$ & $138.84 \pm 4.52$ \\
\hline
\multicolumn{8}{l}{\emph{PDS 70 c}}\\
NACO NB4.05 & $234.80 \pm 13.67$ & $3.46 \pm 28.15$ & $276.79 \pm 1.79$ & $-0.79 \pm 5.25$ & $-0.44 \pm 0.10$ & $238.25 \pm 31.29$ & $275.56 \pm 5.55$ \\
\hline
\end{tabular}
\egroup
\end{sidewaystable}

\section{Posterior distributions}
\label{sec:posterior}

\begin{figure*}
\centering
\includegraphics[width=\linewidth]{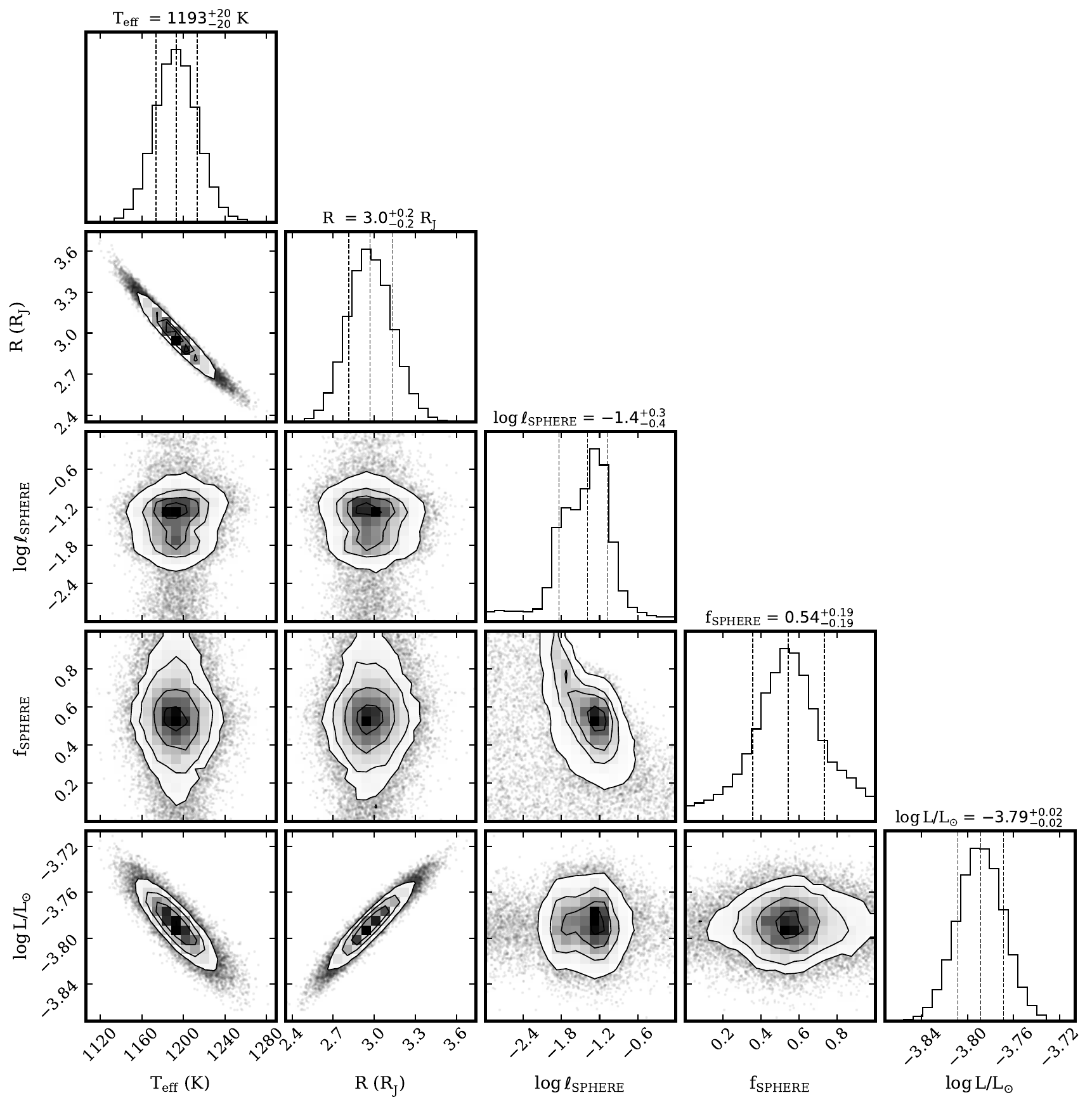}
\caption{Posterior distributions from fitting a single Planck function to the SED of PDS~70~b. The 1D marginalized distributions are shown in the \emph{diagonal} panels and the 2D parameter projections in the \emph{off-axis panels}. The bolometric luminosity, $\log{L/\Lsun}$, has been calculated from the posterior samples of $\Teff$ and $R$.}
\label{fig:posterior_1}
\end{figure*}

\begin{figure*}
\centering
\includegraphics[width=\linewidth]{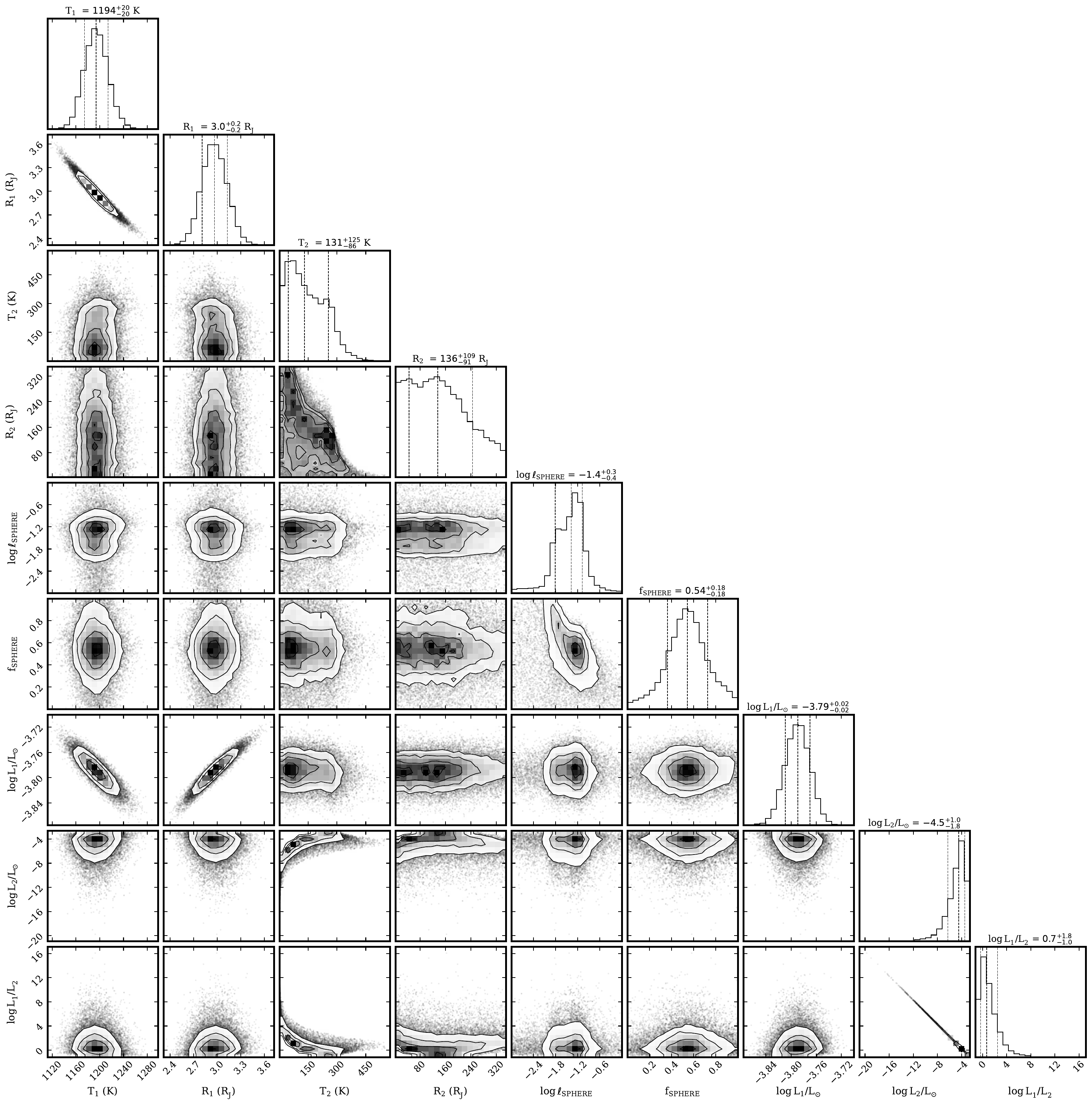}
\caption{Posterior distributions from fitting a combination of two Planck functions to the SED of PDS~70~b. Further details are provided in the caption of Fig.~\ref{fig:posterior_1}.}
\label{fig:posterior_2}
\end{figure*}

Figures~\ref{fig:posterior_1} and \ref{fig:posterior_2} show the 1D and 2D projections of the posterior samples from fitting the photometric and spectroscopic data of PDS~70~b with a model spectrum consisting of one and two blackbody components, respectively. Throughout this work, we have used the median of each parameter as the best-fit value, and the 16th and 84th percentiles as the 1$\sigma$ uncertainties. For the second blackbody component, we have quoted the 84th percentile as the upper limit on the temperature and radius. For a single blackbody component, the fitted (photospheric) temperature and radius are $\Teff$ and $R$, while for two blackbody components, these are given as $T_1$ and $R_1$, $T_2$ and $R_2$ for the first and second component. For the SPHERE/IFS spectrum, we have fitted the logarithm of the correlation length, $\log{\ell_\mathrm{SPHERE}}$ and fractional amplitude of the correlated noise, $f_\mathrm{SPHERE}$ (see Sect.~\ref{sec:modelling_approach} for details).

\end{appendix}

\end{document}